\documentclass{aa}

\newcommand{\mic}{\textmu{}m~} 
\newcommand{\AIBirrad}{$\rm AIB_{\rm Irrad}$}
\newcommand{\AIBshield}{$\rm AIB_{\rm Shield}$}

\usepackage[para]{threeparttable}
\usepackage{natbib}
\usepackage[colorlinks=true, allcolors=blue]{hyperref}
\usepackage{graphicx}
\usepackage{subcaption}         
\usepackage{lscape}             
\usepackage{placeins}           

\usepackage{multirow}
\usepackage{txfonts}
\usepackage{ulem}
\usepackage{xcolor}
\usepackage{colortbl}
\usepackage{xurl}

\begin{document}

   \title{PDRs4All}
   \subtitle{XXIII. JWST-MIRI template spectra of PAHs in the Orion Bar obtained with blind signal separation}

   \author{P. Amiot\inst{1},
           O. Bern\'e\inst{1},
           J. Roux-Morin\inst{1},
           I. Schroetter\inst{1},
           C. Boersma\inst{2},
           A. Fuente\inst{3},
           J. R. Goicoechea\inst{4},
           E. Habart\inst{5},
           A. Maragkoudakis\inst{2},
           T. Onaka\inst{6},
           E. Peeters\inst{7,8},
           M. Robberto\inst{9,10},
           A.G.G.M Tielens\inst{11},
           D. Van De Putte\inst{7,9},
           S. Vicente\inst{12}
          }

   \institute{Institut de Recherche en Astrophysique et Planétologie, Université de Toulouse, Centre National de la Recherche Scientifique (CNRS), Centre National d’Etudes Spatiales, 31028 Toulouse, France
   \and
   NASA Ames Research Center, MS 245-6, Moffett Field, CA 94035-1000, USA
   \and
   Centro de Astrobiología (CAB), INTA-CSIC, Carretera de Ajalvir Km. 4, Torrejón de Ardoz, 28850 Madrid, Spain
   \and
    Instituto de Física Fundamental (CSIC), Calle Serrano 121, 28006, Madrid, Spain
   \and
   Institut d’Astrophysique Spatiale, Université Paris-Saclay, CNRS, Bâtiment 121, F-91405 Orsay Cedex, France
   \and
   Department of Astronomy, Graduate School of Science, The University of Tokyo, 7-3-1 Bunkyo-ku, Tokyo 113-0033, Japan
   \and
    Department of Physics and Astronomy, University of Western Ontario, London, Ontario N6G 2V4, Canada
   \and 
   Institute for Earth and Space Exploration, University of Western Ontario, London, Ontario N6A 5B7, Canada
   \and 
   Space Telescope Science Institute, Baltimore MD, 21212, USA
   \and
   Johns Hopkins University, Baltimore MD, 21212, USA
   \and 
   Astronomy Department, University of Maryland, College Park, MD  20742, USA
   \and
   Instituto de Astrofísica e Ciências do Espaço, Tapada da Ajuda, Edifício Leste, 2$^\circ$ Piso, P-1349-018 Lisboa, Portugal
             }

   \date{Received ?? ; accepted ??}

\titlerunning{JWST-MIRI template spectra of PAHs} 
 \authorrunning{P. Amiot, O. Bern\'e, J. Roux-Morin et al.}

  \abstract
  { 
  The aromatic infrared bands (AIBs) are observed in a wide variety of astrophysical sources, and now in unprecedented detail by the James Webb space telescope (JWST). These bands are generally attributed to the emission from polycyclic aromatic hydrocarbon (PAHs).
Our aim is to extract mid-infrared AIB template spectra from JWST observations of the Orion Bar photo-dissociation region that can be used to derive PAH properties from observations of other objects.
We used the JWST MIRI-MRS observations from the PDRs4All program. We first isolated the AIBs from the data using a multi-component fitting approach. We then applied a non-negative matrix factorization to the isolated AIB spectra and extracted two templates, named \AIBirrad~and \AIBshield~following previous work.
The spatial distribution and spectral characteristics of these two spectra suggest that they originate from two chemical distinct populations. We interpret \AIBirrad~as emission originating from a PAH population strongly exposed to ultra-violet (UV) radiation and mainly composed of large, compact ionized PAHs.
Conversely, \AIBshield~is interpreted as originating from a more shielded population of PAHs, which is less exposed to the UV by being more embedded in the cloud. This population consists of small, asymmetric and mainly neutral PAHs with bonded aliphatic side groups.
These templates can serve to analyze PAH emission in a wide variety of objects observed with MIRI-MRS.
}

   \keywords{}
   \maketitle
%

\section{Introduction}\label{sect:intro}

\begin{figure*}[h]
    \centering
    \includegraphics[width=\linewidth]{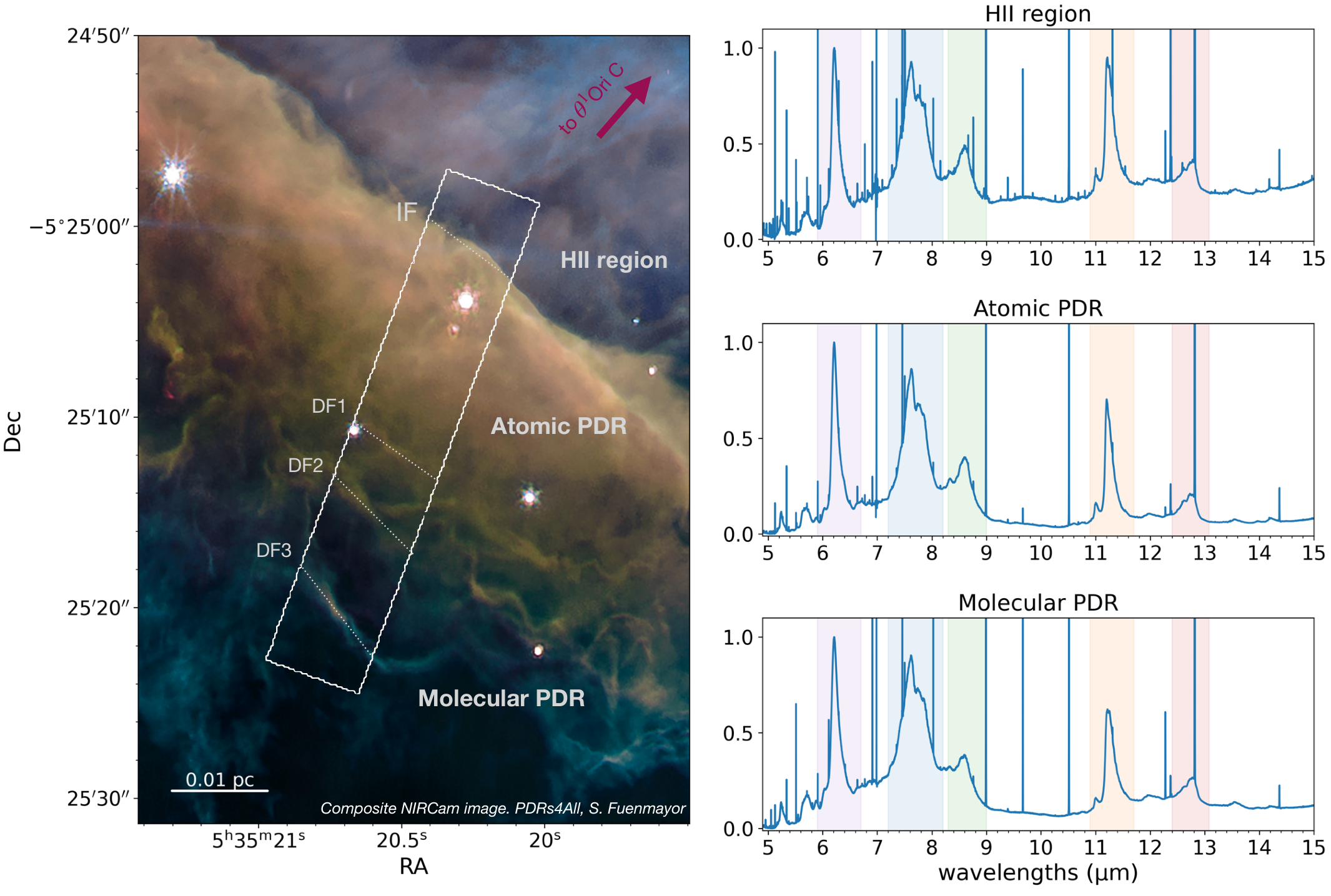}
    \caption{(left) NIRCam composite images of the Orion Bar (see \citealt{habart_pdrs4all_2024} for details on the data, colored image credits S. Fuenmayor). The white dashed box indicates the MIRI-MRS FOV. North is up, east is to the right. (right) MIRI-MRS normalised spectra of the \ion{H}{ii} region, the atomic PDR, and the molecular PDR. The spectra are from \citet{chown_pdrs4all_2024}. The spectrum of the molecular PDR is the mean spectrum of DF1, DF2, and DF3. The colored bands highlight the main PAH bands.}
    \label{fig:NIRCam_MIRI_fov}
\end{figure*}

Polycyclic Aromatic Hydrocarbons (PAHs) are a family of molecules made up of aromatic cycles of carbon and are present in a wide variety of astrophysical sources such as star-forming regions, galaxies, and protoplanetary disks \citep{Leger_Puget1984, Allamandola1989_PAHs, peeters_rich_2002}. 
When ionized, PAHs release electrons with energies of a few eV, which interact with the surrounding gas and raise its temperature \citep{Bakes_Tielens1998}. 
Ionization of PAHs is thought to be the main source of gas heating in most interstellar environments \citep{Berne2022_PAH}.
Moreover, PAHs are efficient absorbers of ultra-violet (UV) photons. After absorption of a UV photon, the electronic excitation is quickly transferred through internal conversion into vibrational excitation modes. Subsequently, PAHs relax through a collection of characteristic stretching and bending modes. These modes produce broad bands in the mid-infrared (IR) dubbed the aromatic infrared bands (AIBs), whose main components are situated at 3.3, 6.2, 7.7, 8.6, 11.2 and 12.7~\textmu{}m.
AIBs dominate the mid-IR spectra of photo-dissociation regions (PDRs), which are regions dominated by far-UV (FUV, $\rm 6<Energy<13.6$ eV) radiation.
These bands show strong variations depending on the observed environment, with band positions shifting and band ratios varying \citep[e.g.][]{Cesarsky2000, Hony_2001, peeters_rich_2002}.
These variations reflect changes in the chemical properties of PAHs, linked to local astrophysical conditions, notably the UV radiation field and atomic and molecular hydrogen densities (see \citealt{Tielens_2026} for a review). 

In order to analyze the connection between astrophysical conditions and AIB shape, various methods enable to decompose AIB spectra, such as PAHFIT \citep{Smith2007_PAHFIT} or PAHdb \citep[e.g.][]{Bauschlicher2010_PAHdb, Rosenberg2011, Boersma2014A_PAHdb, Ricca2026_PAHdb, maragkoudakis_pdrs4all_2026}.
Alternatively, one can use blind signal separation (BSS) methods where the observed variations are decomposed into a set of elementary spectra. The physical interpretation of these elementary spectra can then provides crucial information on the chemical populations within the observed region.
For instance, \citet{Rapacioli2005, Berne_2007, Rosenberg2011} applied BSS to observations of various PDRs with the Infrared Space Observatory (ISO) and Spitzer Space Telescope.  
They extracted three elementary spectra of PAHs, interpreted as the emission from neutral PAHs ($\rm PAH^0$), cationic PAHs ($\rm PAH^+$), and carbonaceous very small grains (VSGs), which were proposed to consist of PAH clusters by \citet{Rapacioli2006}. Following these works, \citet{joblin_2008} proposed a complementary emission component to reproduce the observed AIB emission that arises from the contribution of large ionized PAHs, namely $\rm PAH^x$.
These spectra were then used in the construction of the PAHTAT fitting tool \citep{Pilleri2012_PAHTAT}, which allows the modeling of IR PAH spectra.
Using PAHTAT, \citet{Pilleri2015_eVGS} found that VSGs do not survive in the harsh environments exposed to strong UV radiation, and are only present in the more shielded, embedded parts of PDRs, where UV radiation is weak.
\citet{foschino_learning_2019} provided a fast and robust BSS method, developed in anticipation of the arrival of the James Webb Space Telescope \citep[JWST, ][]{Gardner_2023} and it providing large datasets of medium resolution spectra of AIBs with the integral field units of its MIRI and NIRSpec instruments.
This method was tested on 31 ISO spectra from various astrophysical sources such as reflection nebulae, planetary nebulae, \ion{H}{ii}, and Compact \ion{H}{ii} regions, one post-AGB star and a starburst galaxy. 
This work retrieved four template spectra, namely that for large ionized PAHs, cations, neutrals, and evaporating VSGs. However, this work was limited by the size and heterogeneity of the sample.

Since its arrival in 2022, JWST has observed PAHs in a variety of environments, such as planetary nebulae \citep[e.g.][]{Clark2025}, nearby galaxies \citep[e.g.][]{Egorov2023, Chastenet2023}, distant galaxies \citep[e.g.][]{Spilker_2023} and protoplanetary disks \citep[e.g.][]{berne_far-ultravioletdriven_2024, Sturm2024, Schroetter2025_NatAst, Bandyopadhyay2025_PAH_PPD}. 
This underscores the need for a benchmark analysis of PAH properties and for tools to interpret their behavior in different environments.

One of the most favorable environment to study the properties of PAHs are extended bright PDRs in star-forming regions such as the Orion Bar (Fig.~\ref{fig:NIRCam_MIRI_fov}), located at 390 pc from the Earth \citep{maiz_apellaniz_villafranca_2022}.
The Orion Bar was observed with JWST as part of the Early release science PDRs4All program  \citep[ID: 1288]{berne_pdrs4all_2022}. The detailed analysis of the AIB emission in the Orion Bar has been the subject of many studies part of this program \citep[e.g.,][]{chown_pdrs4all_2024, pasquini_pdrs4all_2024, elyajouri_pdrs4all_2024, khan_pdrs4all_2025, Goicoechea2025_PDRs4All, van_de_putte_pdrs4all_2025_PAHfit, maragkoudakis_pdrs4all_2026, Khan2026_PDRs4All}.
These works highlight the evolution of PAHs throughout the Bar, where the more UV-shielded molecular part is dominated by small PAHs (with a number of carbon $\rm N_C<50$) and a low ionization fraction. Such small PAHs do not survive the neutral atomic region of the PDR that is more exposed to UV radiation. There, the emission is predominately coming from large PAHs ($\rm N_C>70$). The UV-processing of PAHs results in a higher ionization fraction and removes any attached aliphatic groups (such as methyl or ethyl groups). 

However, these works did not provide any AIB template spectra that could be used to interpret other observations. Such templates can be obtained with BSS, which has been extensively applied to various data, but whose application to JWST data remains limited for now. One notable exception is the work of \citet{schroetter_pdrs4all_2024}, who applied BSS to the NIRSpec IFU data of the Orion Bar. They extracted two AIB template spectra that  were then successfully used to study the PAH population in nearby galaxies. This further demonstrates the capabilities of BSS methods to extract physical elementary spectra that can be used to interpret other datasets. 
These template spectra are however limited to the NIRSpec wavelengths range, which only covers the 3.3 and 3.4~\textmu{}m bands and misses most of the AIBs, that are in the 5-15~\textmu{}m wavelength range. 

In this work, we aim to extend the application of the BSS work to the MIRI IFU data on the Orion Bar, and to extract AIB template spectra in the 5-15~\textmu{}m that can be used to interpret AIB spectra observed in various other environments with JWST.

\begin{figure*}[h]
    \centering
    \includegraphics[width=0.95\linewidth]{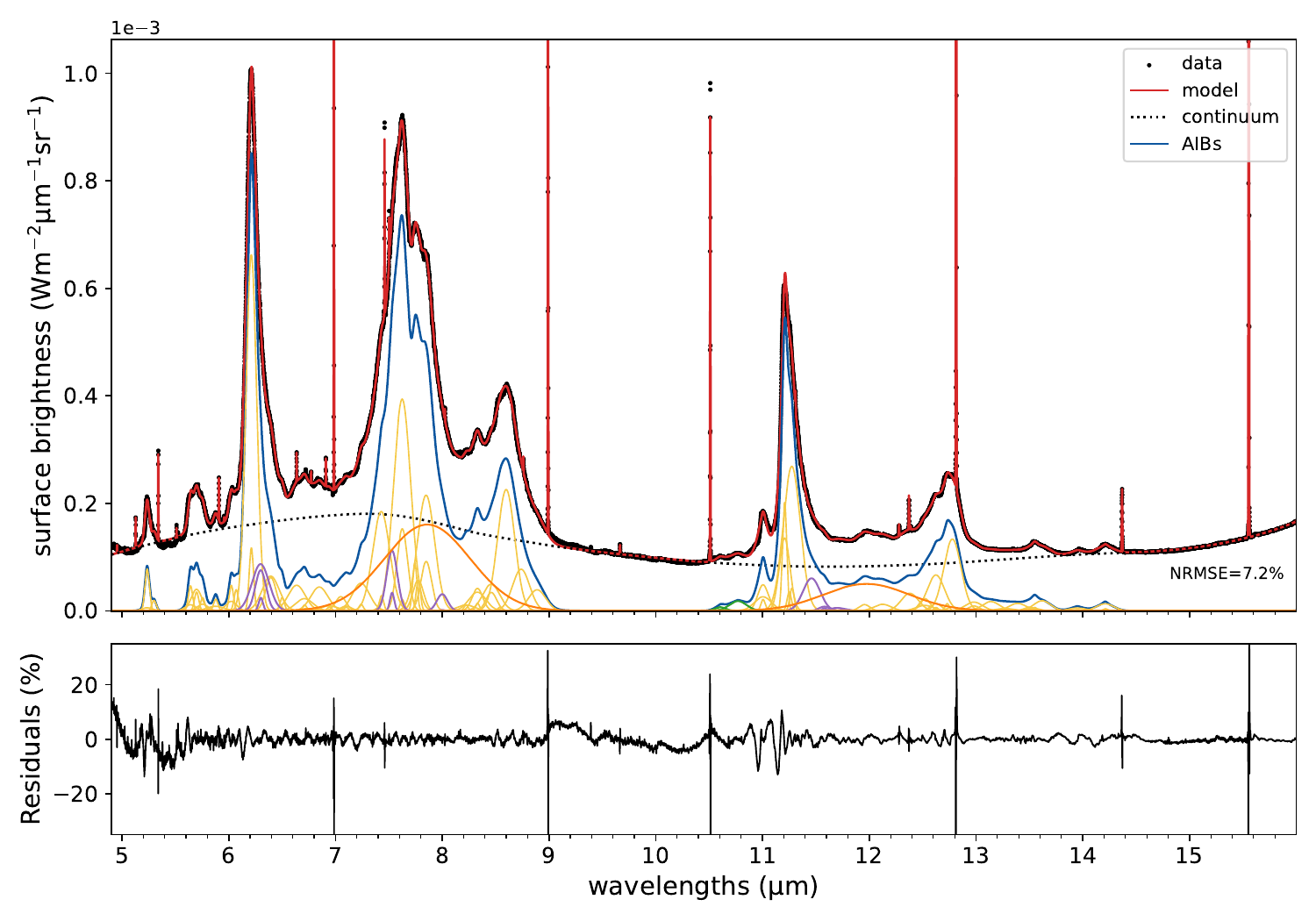}
    \caption{MIRI-MRS spectrum of a region within the atomic PDR (in black). The fitted model is shown with the red line.
    The contribution of the continuum is represented with the dotted black line, and the total PAH contribution (i.e., AIB spectrum) is shown in blue.
    The  orange, purple, yellow and green lines represent the contribution of the subcomponents of the AIB spectrum.
    The yellow lines show the bands reported in \citet{chown_pdrs4all_2024} for the AIBs. The orange lines indicate the additional bands considered to account for the plateau emissions. The green lines represent the bands not reported by Chown et al., and the purple lines represent the bands used to adjust the wings of the 6.3, 7.7, and 11.2~\textmu{}m bands
    The bottom panel indicates the residuals of the fit.}
    \label{fig:gaussian_fit}
\end{figure*}

\section{Observations and data}

The Bar is thought to have formed by compression of the Orion Molecular Cloud, induced by the expansion of the \ion{H}{ii} region produced by the massive stars in the Trapezium cluster \citep{Goicoechea2016_nature}. 
Its nearly edge-on geometry provides the ideal case to spatially resolve the different layers between the \ion{H}{ii} region and the molecular cloud.
The Bar was observed with JWST as part of the PDRs4All Early Release Science program \citep[PID 1288,][]{berne_pdrs4all_2022} with the Near-IR Camera \citep[NIRCam, ][]{Rieke_2023_NIRCam}, the Mid-IR Instrument \citep[MIRI,][]{Argyriou2023_MIRI}, and the Near-IR spectrograph \citep[NIRSpec,][]{Jakobsen2022_NIRSpec}. 
The observations include imaging with NIRCam and MIRI \citep[see][for details]{habart_pdrs4all_2024} as well as mosaics across the Orion Bar obtained with MIRI in its Medium Resolution Spectroscopy (MRS) mode and NIRSpec with its Integral Field Unit (IFU). Observational and data reduction details are provided in \citet{chown_pdrs4all_2024, van_de_putte_pdrs4all_2024} for MIRI-MRS and \citet{peeters_pdrs4all_2024} for NIRSpec-IFU. The field of view (FOV) of the MIRI-MRS observations is shown in Fig.~\ref{fig:NIRCam_MIRI_fov}.

Using the NIRSpec-IFU data, \citet{peeters_pdrs4all_2024} identified five key regions within the Bar, reflecting distinctly different environments. These regions, indicated in Fig.~\ref{fig:NIRCam_MIRI_fov}, are the \ion{H}{ii} region, dominated by ionizing photons and where the hydrogen is mainly ionized; the Atomic PDR (APDR), where the chemistry is controlled by FUV photons and where hydrogen is mainly atomic; and the three dissociation fronts (DF1, DF2, DF3), which separate the atomic region from the molecular region and where hydrogen is fully molecular (H$_2$). 
In fact, the distinction of the three dissociation fronts is a projection effect of the edge-on PDR, which is organized like terraced paddy fields \citep[cf.][]{habart_pdrs4all_2024}. It is also important to note that most of the AIBs (and  $\rm H_2$ lines) observed in the \ion{H}{ii} region actually arise from the face-on background PDR at the rim of OMC-1, since PAHs are not expected to survive in this region.

We use the MIRI-MRS observations as reduced by \citet{chown_pdrs4all_2024}.
We use the fully stitched cube, which contains the 12 sub-bands reprojected onto the spatial grid of the short wavelength subchannel channel 1 observations. Details on the classical stitching method used for these cubes are provided by \citet{canin_2026_haute_couture}.
The fully stitched cube thus covers the short-medium-long subchannels of the four MIRI channels, spanning a wavelength range from 4.9 to 28.7~\textmu{}m, with a pixel size of 0.207\textquotedbl/pix. The spectral resolution varies from $\sim$1500--3500 across the wavelength range. 
We focus on the 5--15~\textmu{}m region, which covers the main AIBs at 6.2, 7.7, 8.6, 11.2 and 12.7~\textmu{}m. This reduces the final spectral dimension of the cube to 27,522 spectral points\footnote{The spectral dimension of each channel was oversampled to ensure a common wavelength grid. Therefore, the number of spectral points does not reflect the actual spectra resolution.}.
The final data cube, $C(i,j,\lambda)$ (with $i, j$ covering the spatial dimension and $\lambda$ the spectral dimension), has dimensions $(n_i,n_j,m_\lambda)$ with  $n_i =41$, $n_j=213$ for the two spatial dimensions and $m_\lambda=27,522$ for the spectral dimension.

\section{Method: extraction of AIB template spectra}\label{sect:Method}

We aim to extract MIRI template spectra of AIBs, following the method used by \citet{schroetter_pdrs4all_2024} for the NIRSpec-IFU data. 
First, we build the data matrix $\cal{M}$ with initial dimensions $(n_i\times n_j, m_\lambda)$, whose rows correspond to the MIRI-MRS spectra. These spectra are computed as the median spectra extracted from $C(i,j,\lambda)$ within a cross of five spatial pixels (one central pixel and the top/bottom and left/right pixels). This ensures coverage of the wavelength-dependent point spread function. Spectra situated at the edges of the FOV are excluded. 
Also excluded from $\cal{M}$ are spectra with more than 100 (>0.36\%) bad points (i.e., zeros and NaNs). This generally corresponds to spatial pixels close to the edges of the FOV. The final data matrix $\cal{M}$ contains 7,314 spectra, each with 21,147 points, corresponding to all the clean 5--15~\textmu{}m spectra.
We describe in section~\ref{sect:AIBextraction} the method used to isolate the AIB from $\cal{M}$. In section~\ref{sect:NMF}, we present the extraction method of the AIB template spectra.

\subsection{AIB extraction}\label{sect:AIBextraction}

Fig.~\ref{fig:gaussian_fit} shows an example of a MIRI-MRS spectrum extracted from five spatial pixels in the atomic PDR. The observed spectrum contains multiple features, such as atomic and molecular emission lines (mainly \ion{H}{i} and rovibrational $\rm H_2$ lines), AIBs, and a continuum emission from stochastically heated small dust grains. To isolate the AIBs, the first step is to separate them from other spectral contributions. 
To do so, we follow the linearized method of \citet{foschino_learning_2019}. This method uses a multicomponent model to decompose the contribution from atomic emission lines, the continuum emission from small dust grains, and AIBs. 
We do not include the effect of dust extinction along the line of sight for AIBs and atomic emission lines. This is motivated by the fact that the extinction is expected to be negligible at the MIR. Moreover, \citet{van_de_putte_pdrs4all_2025_PAHfit} found that PAH emission from the Orion Bar is unattenuated, i.e., $\rm A_V\sim 0$. 
The atomic lines are modeled as Gaussian profiles with standard deviation $\sigma$ equal to the spectral resolution and central wavelengths given in the line list from \citet{van_de_putte_pdrs4all_2024}. The continuum emission is fitted over the 5-15~\textmu{}m range using a series of modified black bodies, considering the extinction curve of \citet{Weingartner_Draine_2001} with $R_\text{V} = 5.5$, at temperatures between 40 and 6,000~K with a step size of 20~K. 
For our analysis, the AIBs are also fitted with Gaussian profiles, centered at the wavelengths presented in the catalog of AIBs by \citet{chown_pdrs4all_2024} (cf. Table~\ref{tab:inv}). We consider a full width at half maximum (FWHM) of 0.02, 0.05, and 0.08~\textmu{}m for these Gaussian profiles.
This catalog is augmented with Gaussian profiles at 10.6 and 10.77~\textmu{}m, with the same FWHM as before, to account for the AIBs at 10.6 and 10.8~\textmu{}m \citep{Hony_2001, khan_pdrs4all_2025}. 
We also add Gaussian profiles at 6.3, 7.53, 8, 11.46, 11.58, and 11.70~\textmu{}m to better model the wings observed for certain AIBs \citep[cf.][Section 3]{chown_pdrs4all_2024}. Finally, we consider two Gaussian profiles centered at 7.9 and 12~\textmu{}m, each with a width of 0.4~\textmu{}m, to account for plateau emission.

Each spectrum in $\cal{M}$ is fitted with the model using a non-negative least square method. Fig.~\ref{fig:gaussian_fit} shows an example of a fitted spectrum, with each contributing Gaussian representing AIB shown separately. 
The residuals of the fits are generally below 10\%, except for the atomic lines, which can be off by as much as 50\%. 
The total errors of the fits are estimated considering the normalized root mean square error (NRMSE) defined as in \citet{boulais_unmixing_2021}, which corresponds to the norm of the deviation between the data and the model, normalized to the norm of the data. 
Keeping only the contribution of AIBs (blue curve Fig.~\ref{fig:gaussian_fit}), we build the ``pure-AIB'' signal matrix, $\textbf{X}_{\rm AIB}$. The dimension of this matrix is 7,314$\times$21,147 (the same as $\cal{M}$), with the 7,314 rows corresponding to the isolated AIB spectra, each with 21,147 spectral points.
We note that while the model reproduces the data with a good fidelity, we still lose information during this process as indicated by the residuals observed Fig.~\ref{fig:gaussian_fit}. The errors in the fits are due to both the intrinsic noise within the data, and the accuracy of the model. 
This implies for the next that, at best, the BSS (section~\ref{sect:NMF}) will be at this level of precision. 

The applied method to fit the AIBs is comparable to that used by PAHFIT\footnote{\url{https://github.com/pahfit/pahfit/}} \citep{van_de_putte_pdrs4all_2025_PAHfit}. PAHFIT enables a detailed analysis of AIBs positions, strengths, and widths. It fits the AIBs, thermal dust emission, stellar light, and gas emission lines considering the effect of dust extinction. The AIBs are reproduced with a collection of Drude profiles, whose positions, widths and heights are adjusted to fully characterize each subcomponent.
While this provides a comprehensive characterization of the AIBs and other spectral components, it comes at higher computational cost than linear methods.

\subsection{MASS-NMF decomposition}\label{sect:NMF}

\begin{figure*}
    \centering
    \includegraphics[width=0.9\linewidth]{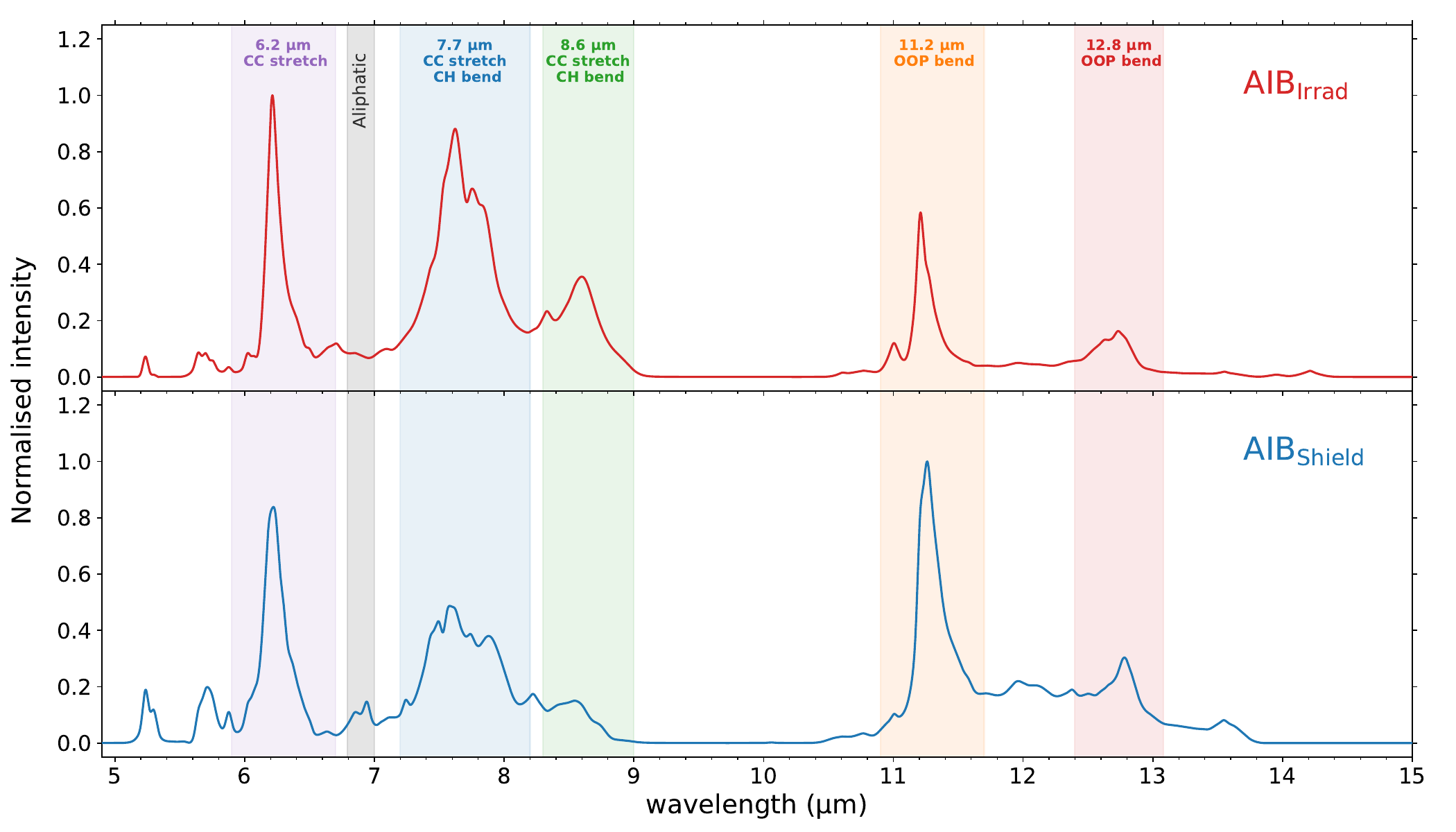}
    \caption{Elementary spectra obtained applying MASS-NMF with $r=2$. $\rm AIB_{\rm Irrad}$ is shown in red and $\rm AIB_{\rm Shield}$ in blue. 
    The spectra are normalized to their peak intensity and in units compatible with $\rm W~m^{-2}$\textmu{}m$^{-1}$.}
    \label{fig:MIRItemplates}
\end{figure*}

We apply a non-negative matrix factorization (NMF) method on $\textbf{X}_{\rm AIB}$, following the same approach as \citet{Berne_2007, foschino_learning_2019, schroetter_pdrs4all_2024}.
Briefly, this method enables the decomposition of a non-negative signal into a mixture of $r$ source signals, corresponding to AIB template spectra. 
It allows expressing $\textbf{X}_{\rm AIB}$, of dimension $n\times m$, as the product of a weight matrix \textbf{W}, with dimension $n\times r$, and a source matrix \textbf{S} with dimension $r\times m$. The $r$ rows of \textbf{S} are the sought AIB template spectra. We consider $r=2$, following \citet{schroetter_pdrs4all_2024}. 
This choice is further discussed in Section \ref{sect:Validation}, and the viability of considering $r>2$ is discussed in Appendix~\ref{sect:NMF_test_r}. 
We used the \texttt{NMF} implementation from the \textsc{sklearn} Python library. The NMF is initialized with the Maximum Angle Source Separation (MASS) method described in \citet{boulais_unmixing_2021} and a random weight matrix \textbf{W}. 
We used 1,000 different initializations with different \textbf{W}s and considered the mean of the obtained matrices \textbf{S}. The standard deviation estimated on these 1,000 iterations is below 1\%, demonstrating that the NMF is stable against initialization for $r=2$.
The two elementary spectra obtained from the mean \textbf{S} are shown in Fig.~\ref{fig:MIRItemplates}.

\section{Results}\label{sect:results}

Fig.~\ref{fig:MIRItemplates} shows the two templates extracted from the MASS-NMF method. In this section, we first discuss the validity of the results and then describe the spectral characteristics of the template spectra, which we call \AIBirrad~and \AIBshield~in link with the local physical conditions where these populations are present as in \citet{schroetter_pdrs4all_2024}.

\subsection{Validity of the MASS-NMF results}\label{sect:Validation}
\begin{figure*}
    \centering
    \includegraphics[width=0.9\linewidth]{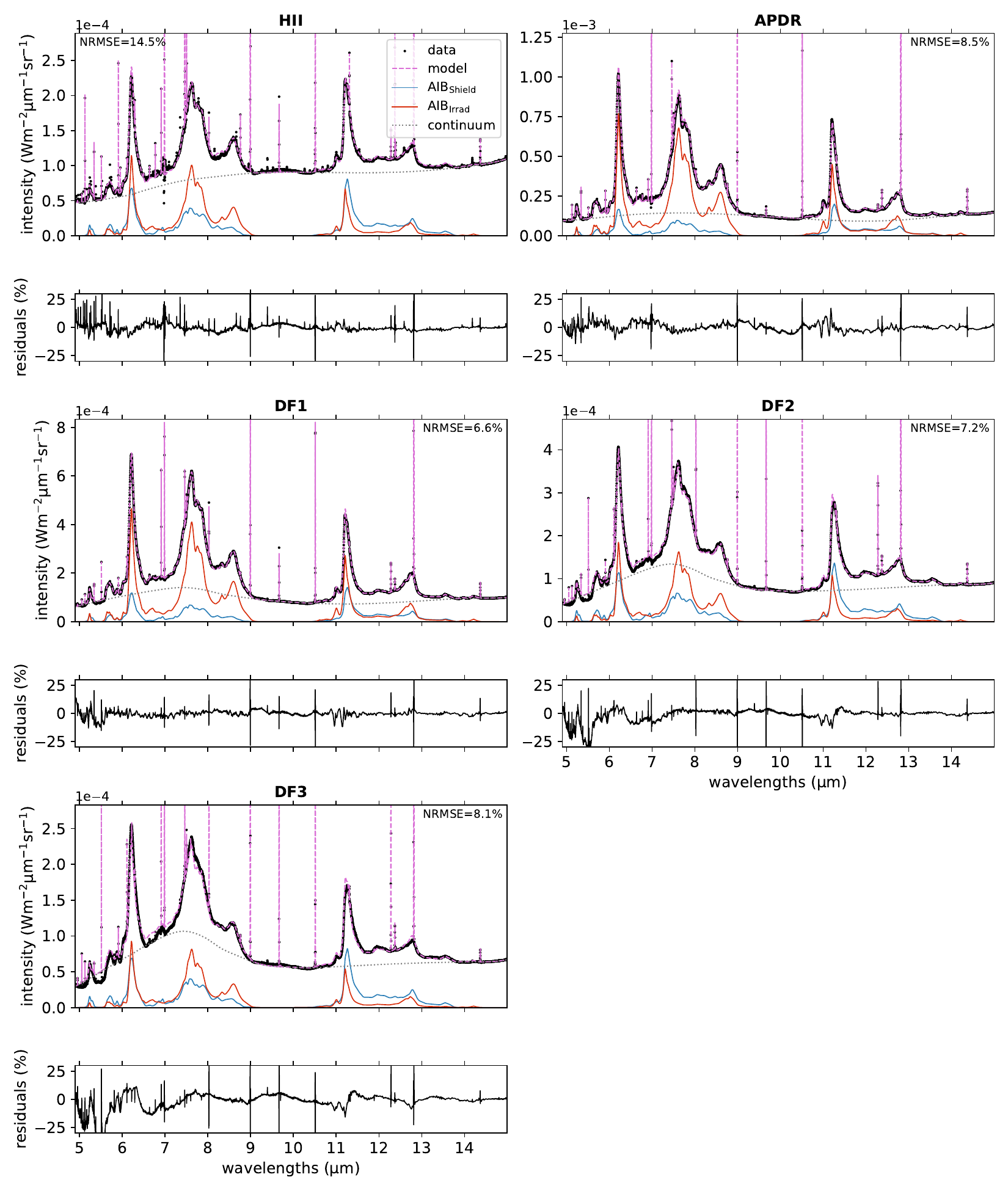}
    \caption{Fit to the spectra of the five key regions identified in the Orion Bar by \citet{peeters_pdrs4all_2024, chown_pdrs4all_2024}, and their associated residuals.
    The five regions are the \ion{H}{ii} region (HII), the atomic PDR (APDR), and the dissociation fronts 1, 2, and 3 (DF1, DF2, DF3). The data are shown in black. The best-fit model is indicated in dashed purple, the continuum emission in dotted grey, and the contribution of \AIBirrad~and \AIBshield~in red and blue, respectively.
    }
    \label{fig:chown_fit}
\end{figure*}

In order to assess the validity of the extracted spectra (Fig.~\ref{fig:MIRItemplates}), we use the \AIBirrad ~and \AIBshield~ templates to reproduce the observed spectra in the five key regions identified in the Orion Bar presented in \citet{chown_pdrs4all_2024}, namely the \ion{H}{ii} region, the atomic PDR (APDR), and the three dissociation fronts DF1, DF2 and DF3.  
To conduct this test, we used the fitting approach described in Section~\ref{sect:AIBextraction}, except we use the two template spectra to model the AIBs (instead of the catalog of Gaussian profiles). 
Fig.~\ref{fig:chown_fit} shows the results of this procedure. 
The global quality of the fits is good. 
The NRMSE is $\le$9\% for the spectra of the atomic PDR, DF1, DF2, and DF3. It is about 14.5\% for the \ion{H}{ii} region spectrum. In that latter case, the error primarily arises from the multitude of atomic lines that are not well reproduced by the model (see Fig.~\ref{fig:chown_fit}, top panel).
Masking the main \ion{H}{i} lines, the NRMSE falls below 4\% for all five spectra, demonstrating that the model reproduces the continuum and the AIBs well. It should be noted that in all five spectra, there is always a contribution from \AIBirrad~and \AIBshield. This demonstrates that the observationally extracted templates of Chown et al. are a mixture of the recovered template spectra. 

The good quality of the reconstruction indicates that using $r=2$ for the NMF (cf. section~\ref{sect:NMF}) provides satisfactory results hence that $r>2$ is not necessary.
 In fact, in Appendix~\ref{sect:NMF_test_r}, we show that considering $r>2$ introduces significant uncertainty on the extracted spectra while it does not significantly decrease the NRMSE.

\subsection{Spatial distribution of \AIBirrad~and \AIBshield }\label{sect:spatial_distribution}

\begin{figure*}[h]
    \centering
    \includegraphics[width=0.85\linewidth]{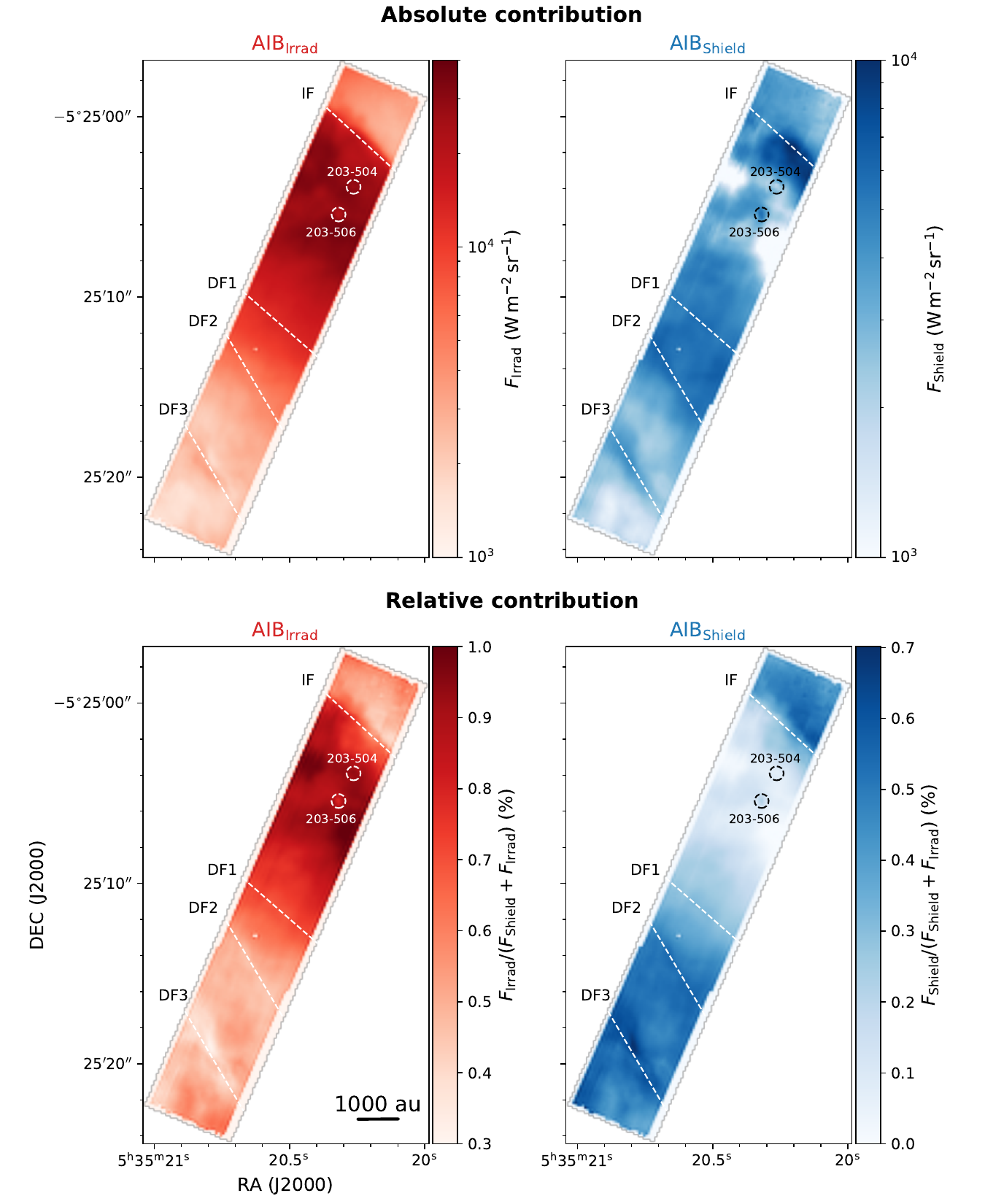}
    \caption{(Top) Absolute radiance of  \AIBirrad, $F_{\rm Irrad}$ (left) and of \AIBshield, $F_{\rm Shield}$ (right) across the full MIRI-MRS FOV.
    (Bottom) Relative contribution of  \AIBirrad~(left) and \AIBshield~(right). The IF, DF1, DF2, and DF3 are indicated with the white dashed lines. The positions of the irradiated protoplanetary disks d203-504 and d203-506 are indicated with the dashed circles. }
    
    \label{fig:MIRImap}
\end{figure*}

Following the fitting approach described above, we fitted the 7,314 spectra from $\cal{M}$ with the two template spectra of Fig.~\ref{fig:MIRItemplates}. 
Fig.~\ref{fig:MIRImap} presents the maps of the absolute and relative spatial contribution\footnote{This corresponds to the radiance of the spectra, i.e the intensity integrated along the spectral axis.} of each template spectrum. 
These maps clearly indicate that one template spectrum dominates the emission in the atomic PDR, between the IF and DF1. It contributes more than $80\%$ to the total emission in this region. The other template spectrum mostly contributes to the emission in the molecular PDR and peaks at the position of DF3, and to a lesser extent at DF2.
The \ion{H}{ii} region in Fig.~\ref{fig:MIRImap} is an almost-equal contribution of \AIBshield~and \AIBirrad.
However, as mentioned earlier,  the PAH emission in this region arises from the face-on background PDR at the rim of OMC-1, and does not trace the \ion{H}{ii} region itself, where PAHs are expected to be destroyed \citep[see][]{peeters_pdrs4all_2024}. 

The relative spatial contributions of the two MIRI templates follow that observed by \citet{schroetter_pdrs4all_2024} for their NIRSpec templates, dubbed \AIBirrad~and \AIBshield.
As mentioned at the beginning of Sect.~\ref{sect:results}, we have therefore adopted the same terminology for the two MIRI templates spectra extracted using BSS here. 
The physical and chemical meaning of these template spectra will be established in Section~\ref{sect:chemical_prop}.

\subsection{Spectral properties of \AIBirrad~and \AIBshield}\label{sect:spectral_prop}

The two extracted spectra, \AIBshield~and \AIBirrad (Fig.~\ref{fig:MIRItemplates}), are dominated by the strong bands at 6.2, 7.7, 11.2~\textmu{}m and, to a lesser extent, by the bands at 8.6 and 12.8~\textmu{}m. 
The 6.2~\textmu{}m band is known to trace aromatic C-C stretching modes, and the 7.7~\textmu{}m complex (encompassing the 7.7~\textmu{}m and 8.6~\textmu{}m bands) is associated with a combination of C-C stretching modes and C-H in-plane bending modes \citep{Allamandola1989_PAHs}. \AIBirrad~is dominated by the 6.2 and 7.7~\textmu{}m  bands. Conversely, \AIBshield~is dominated by the 11.2~\textmu{}m band, which arises from a mixture of {out-of-plane (OOP)} bending modes of solo, duo or trio hydrogens.
Both spectra also exhibit weak sub-features, forming complexes with different shapes. This is particularly true for $\rm AIB_{\rm Shield}$, which has a large collection of these weak bands. The analysis of these bands, supported by Fig.~\ref{fig:template_zoom}, is detailed in the following: 

\begin{figure}
    \centering
    \includegraphics[width=\linewidth]{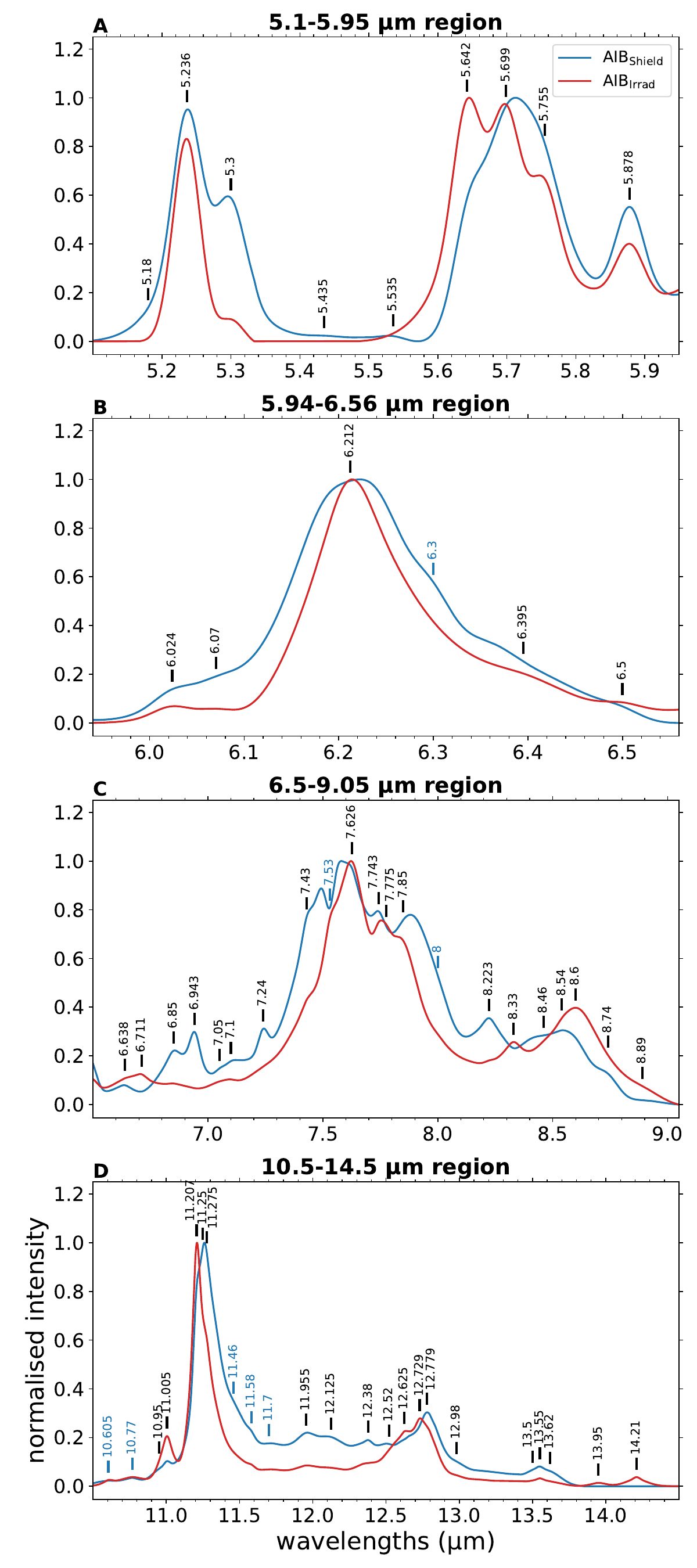}
    \caption{Template spectra \AIBshield~(in blue) and \AIBirrad~(in red) in the 5.1--5.95 (\textbf{A}), 5.94--6.56 (\textbf{B}), 6.5--9.05 (\textbf{C}), and 10.5--14.5~\textmu{}m region (\textbf{D}), from top to bottom. The spectra are normalized to their local maximum intensity in each region, in units of $\rm W~m^{-2}~\mu m^{-1}$. The vertical black lines mark the bands reported in \citet{chown_pdrs4all_2024}, and the blue vertical lines mark the position of the Gaussian profiles added to reproduce the wings of the bands (cf. Section~\ref{sect:AIBextraction}).
    }
    \label{fig:template_zoom}
\end{figure}

\begin{itemize}
    \item \textbf{5--6~\textmu{}m range:} 
    This spectral range is dominated by the 5.25~\textmu{}m and the 5.75~\textmu{}m bands, which are combination bands generated by OOP modes \citep{Boersma2009, Vats2026, White2026}. Fig.~\ref{fig:template_zoom}\textbf{A} shows the normalized spectra in this wavelength range. The peak to peak 5.25/5.75 ratio is slightly higher in \AIBshield.
    The 5.25~\textmu{}m band is composed of three subcomponents, located at 5.18, 5.236, and 5.30~\textmu{}m. In both spectra, the band is dominated by the 5.236~\textmu{}m subfeature. It is broader in \AIBshield. The 5.30/5.236 ratio is greater for \AIBshield.    
    The 5.18~\textmu{}m subfeature is not present in any of the spectra, even though it is included in the catalog of Gaussian used for the preprocessing (cf. section~\ref{sect:AIBextraction}). This likely originates from a lack of precision of the fit for this band in the preprocessing, and illustrates the dependence of the final template shapes on the AIB fitting procedure.
    
    The 5.75~\textmu{}m band is composed of three subcomponents, located at 5.642, 5.699, and 5.755~\textmu{}m. 
    \citet{van_de_putte_pdrs4all_2025_PAHfit} observed that the bluest of these subcomponents dominates the 5.75~\textmu{}m band in the atomic PDR, while the reddest subcomponent dominates the band in the spectrum of DF2 and DF3. 
    All of these subcomponents are detected in \AIBirrad, with the 5.642 being the strongest.   
    In \AIBshield, the components are less clearly distinguishable. The 5.75~\textmu{}m feature appears as a single broad band peaking at 5.7~\textmu{}m, primarily made up of the 5.699, and 5.755~\textmu{}m subcomponents. {The 5.642 band is almost not visible, except for a weak shoulder presents in the 5.75~\textmu{}m feature}.  
    The 5.878~\textmu{}m feature is clearly detected in both spectra, with a 5.878/5.75 ratio stronger in \AIBshield.    

    \item \textbf{6.2~\textmu{}m band:} The 6.2~\textmu{}m band, which traces the pure aromatic C-C stretch, peaks at 6.212~\textmu{}m in both spectra (cf. Fig.~\ref{fig:template_zoom}\textbf{B}). However, the band is broader in \AIBshield, with the 6.07 and 6.395~\textmu{}m subfeatures clearly detected.
    The 6.024~\textmu{}m feature, which likely arises from C=O stretching modes in {carbonyl functionalized aromatic molecules} \citep[e.g.][]{Sarre2019}, is present in both spectra, but the 6.024/6.2 ratio is larger for \AIBshield.
    The strength of the 6.2~\textmu{}m band is known to be sensitive to the charge state, i.e., to an increase of cations. 
    \item \textbf{aliphatic bands:} 
    The 6.850 and 6.943~\textmu{}m bands (Fig.~\ref{fig:template_zoom}\textbf{C}) likely trace the C-H bending modes in aliphatic groups. Alternatively, they can originate from C-C modes in specific PAHs. Both of them clearly peak in \AIBshield, which covers DF2 and DF3. These regions shows the strongest 3.4~\textmu{}m aliphatic band \citep{peeters_pdrs4all_2024}, such that we interpret the 6.850 and 6.943~\textmu{}m bands as aliphatic bands.
    \item \textbf{7.7~\textmu{}m band:} The 7.7~\textmu{}m band, shown in Fig.~\ref{fig:template_zoom}\textbf{C}, is produced by a blend of aromatic C-C stretching and C-H in plane bending modes. The feature is made up of a collection of subcomponents whose positions are given Table~\ref{tab:inv}. From a general point of view, the 7.7~\textmu{}m band appears more complex in \AIBshield~than in \AIBirrad. The band in \AIBirrad~is primarily composed of the 7.626, 7.775, and 7.85~\textmu{}m subcomponents and peaks at 7.626~\textmu{}m. 
    In \AIBshield, it peaks at 7.58~\textmu{}m. 
    The major components are at 7.626, 7.743, and 7.85~\textmu{}m. A new component, which was not in the initial catalog of Gaussian in the preprocessing, is detected at 7.495~\textmu{}m. The overall 7.7~\textmu{}m then appears broader in \AIBshield. 
    The intermediate wavelength range between the 7.7~\textmu{}m and 8.6~\textmu{}m bands contains the two features at 8.223 and 8.33~\textmu{}m. The first one appears in \AIBshield, whereas the second in \AIBirrad.
    
    \item \textbf{8.6~\textmu{}m band}  The 8.6~\textmu{}m band (Fig.~\ref{fig:template_zoom}\textbf{C}) is made up of many subcomponents in \AIBshield, located at 8.46, 8.54, 8.6, 8.74, and 8.89~\textmu{}m. It peaks is at 8.54~\textmu{}m.
    In \AIBirrad, it appears as a single broad band at 8.6~\textmu{}m, with a red wing extending with the 8.74 and 8.89~\textmu{}m features. 
    The 8.6~\textmu{}m band is associated to a mixture of C-H in plane bending and C-C stretching modes, similarly to the 7.7~\textmu{}m band. It has been suggested that an increase in strength and a shift of the peak to longer wavelengths is connected to an increase in PAH size \citep{Bauschlicher2008_largePAHs}.

    \item \textbf{10-15~\textmu{}m range:} The bands in the 10--15~\textmu{}m range, shown in Fig.~\ref{fig:template_zoom}\textbf{D}, are produced by C-H OOP bending modes. The 11.005 and 11.2~\textmu{}m band are primarily due to OOP modes of solo H \citep{Hony_2001}.
    The 11.2~\textmu{}m band is made up of three subcomponents, located at 11.207, 11.25, and 11.275~\textmu{}m \citep{khan_pdrs4all_2025}. 
    The 11.25 and 11.275~\textmu{}m appear as a single broad band in both template spectra. 
    In \AIBshield, the band is broader and peaks at 11.25~\textmu{}m, while it appears narrow and peaks at 11.207~\textmu{}m in \AIBirrad.    
    The 11.005/11.2 ratio is larger for \AIBirrad.      
    
    The 11.955 and 12.145~\textmu{}m features, which arise from C-H OOP modes of duo H, are both primarily detected in \AIBshield. 
    The 12.8~\textmu{}m band, arising from OOP modes of duo and trio Hs, is dominated by the 12.625, 12.729, and 12.779~\textmu{}m subcomponents. 
    The band peaks at 12.779~\textmu{}m in \AIBshield, while it peaks at 12.729~\textmu{}m in \AIBirrad, where it appears broader. 
    The subcomponents at 13.5, 13.55, and 13.62~\textmu{}m, tracing quartet C-H OOP bend, are all detected in \AIBshield. The 13.55~\textmu{}m subcomponent is also detected in \AIBirrad, though much weaker with respect to the 12.8~\textmu{}m band.
    The 13.95 and 14.21~\textmu{}m features are exclusively present in \AIBirrad. These bands likely arise from OOP modes of quintet H \citep{Hony_2001}, or from CCC skeletal modes in large PAHs \citep{Ricca2012}. Given the absence of the quartet modes in \AIBirrad, we consider the latter origin for the 13.95 and 14.21~\textmu{}m bands. Moreover, it is unlikely to have quintet hydrogens, which are prone to photodestruction, in cationic species \citep{khan_pdrs4all_2025}.

    \item  \textbf{Band ratios:}  
    The 6.2~\textmu{m} band and the 7.7~\textmu{}m complex appear stronger relatively to the 11.2~\textmu{}m band in \AIBirrad~than in \AIBshield~(cf. Fig.~\ref{fig:MIRItemplates}). 
    The 6.2/11.2, 7.7/11.2, and 8.6/11.2 peak to peak ratios are 1.7, 1.5, and 0.6 in \AIBirrad, respectively.
    In \AIBshield, these ratios are 0.8,  0.5, and 0.2. 
    These ratios are known to increase with the ionization fraction of PAHs \citep[e.g.][]{Hudgins1995, Galliano2008, Boersma2015, maragkoudakis_pdrs4all_2026}, even though this interpretation can be biased by PAH size \citep[e.g.][]{Khan2026_PDRs4All}.
    The 6.2/8.6 and 7.7/8.6 ratio are a possible tracer of the medium-to-large PAHs (\citealt{Khan2026_PDRs4All}, Schefter et al. in prep.). The 6.2/8.6 is 2.8 in \AIBirrad~and 5.6 in \AIBshield. The 7.7/8.6 ratio is 2.5 in \AIBirrad~and 3.2 in \AIBshield.
    These band ratios are summarized Table~\ref{tab:ratios}.

\end{itemize}

\section{Comparison with earlier studies}\label{sect:previous_studies}

\subsection{Comparison to spectra obtained with previous blind signal separation studies}

\begin{figure}[h]
    \centering
    \includegraphics[width=\linewidth]{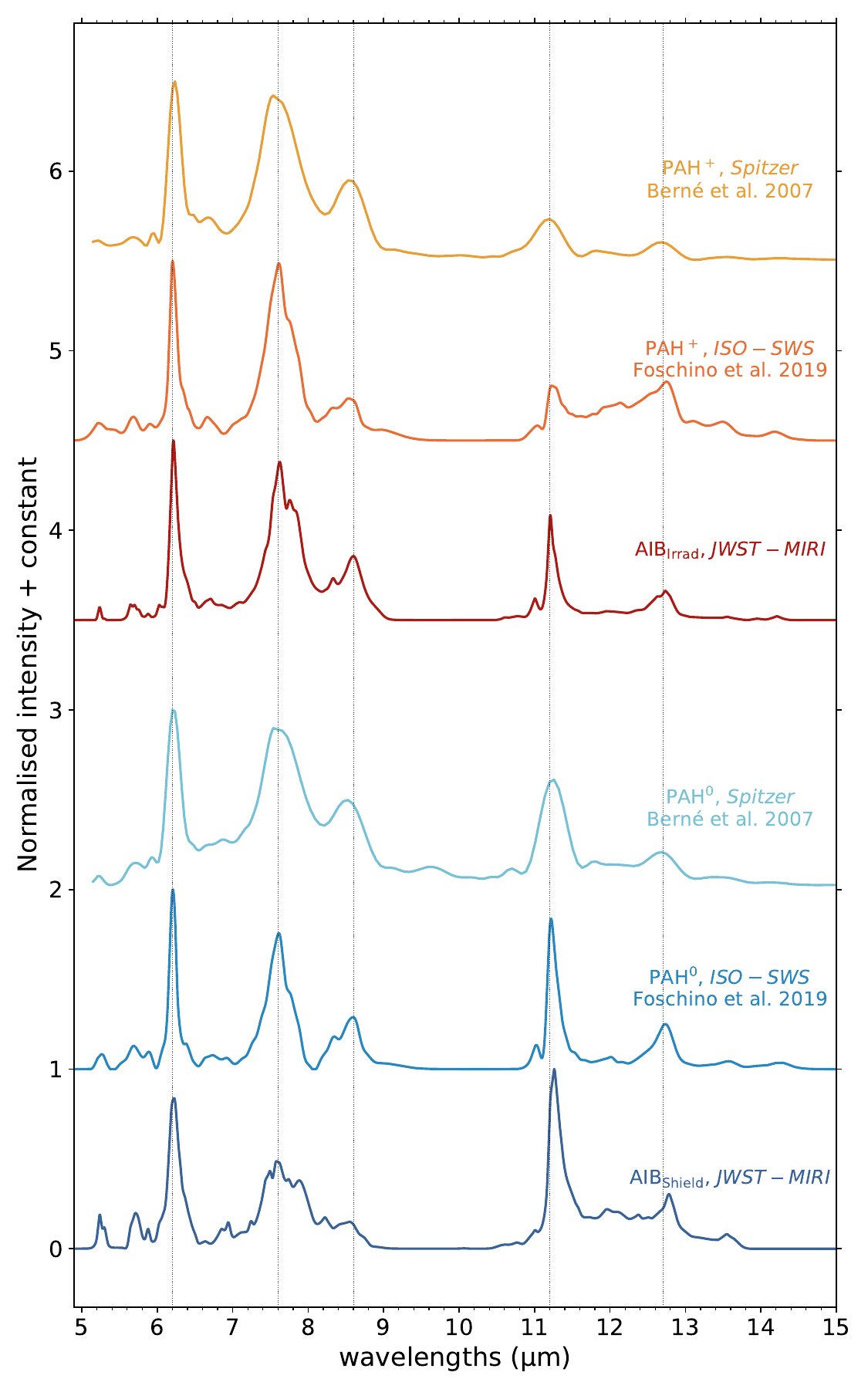}
    \caption{Elementary PAH spectra of cations, $\rm PAH^+$ and neutrals, $\rm PAH^0$ extracted from Spitzer observations of NGC~7023-NW by \citet{Berne_2007}, and from ISO observations of different environments by \citet{foschino_learning_2019} together with \AIBshield~and \AIBshield~from this work. 
    The spectra are normalized to their peak intensity, and shifted for clarity. Units are in $\rm W~m^{-2}$\textmu{}m$^{-1}$. The vertical dotted lines are placed at 6.2, 7.6, 8.6, 11.2 and 12.7~\textmu{}m.}
    \label{fig:BSS_comparison}
\end{figure}

Fig.~\ref{fig:BSS_comparison} shows the $\rm PAH^+$ and $\rm PAH^0$ template spectra extracted by \citet{Berne_2007} from observations of NGC~7023-NW and by \citet{foschino_learning_2019} from various observations (cf. Section~\ref{sect:intro}), as well as \AIBirrad~and \AIBshield~from this work. 
One striking difference is the significant improvement in spectral resolution in \AIBirrad~and \AIBshield~with respect to the other spectra. This improvement demonstrates the capability of JWST to capture the subcomponents of AIBs with unprecedented details. 

\AIBirrad~presents common characteristics with the $\rm PAH^+$ spectra from \citet{Berne_2007} and \citet{foschino_learning_2019}, with strong 6.2 and 7.7~\textmu{}m bands with respect to the 11.2~\textmu{}m. Conversely, the spectral characteristics of the $\rm PAH^0$ spectra resemble that of \AIBshield, with a strong 11.2~\textmu{}m band with respect to the 6.2 and 7.7~\textmu{}m bands, and a larger 11.2/12.8 ratio than $\rm PAH^+$ and \AIBirrad. 
This suggests a similar nature for the carriers of \AIBirrad~and $\rm PAH^+$, as well as \AIBshield~and $\rm PAH^0$. However, it is important to note that the dataset used in \citet{Berne_2007} and \citet{foschino_learning_2019} includes data from regions with conditions that strongly differ from the Orion Bar. 
Thus, the link between the spectra and the carriers should be carefully considered. 

\begin{figure}
    \centering
    \includegraphics[width=\linewidth]{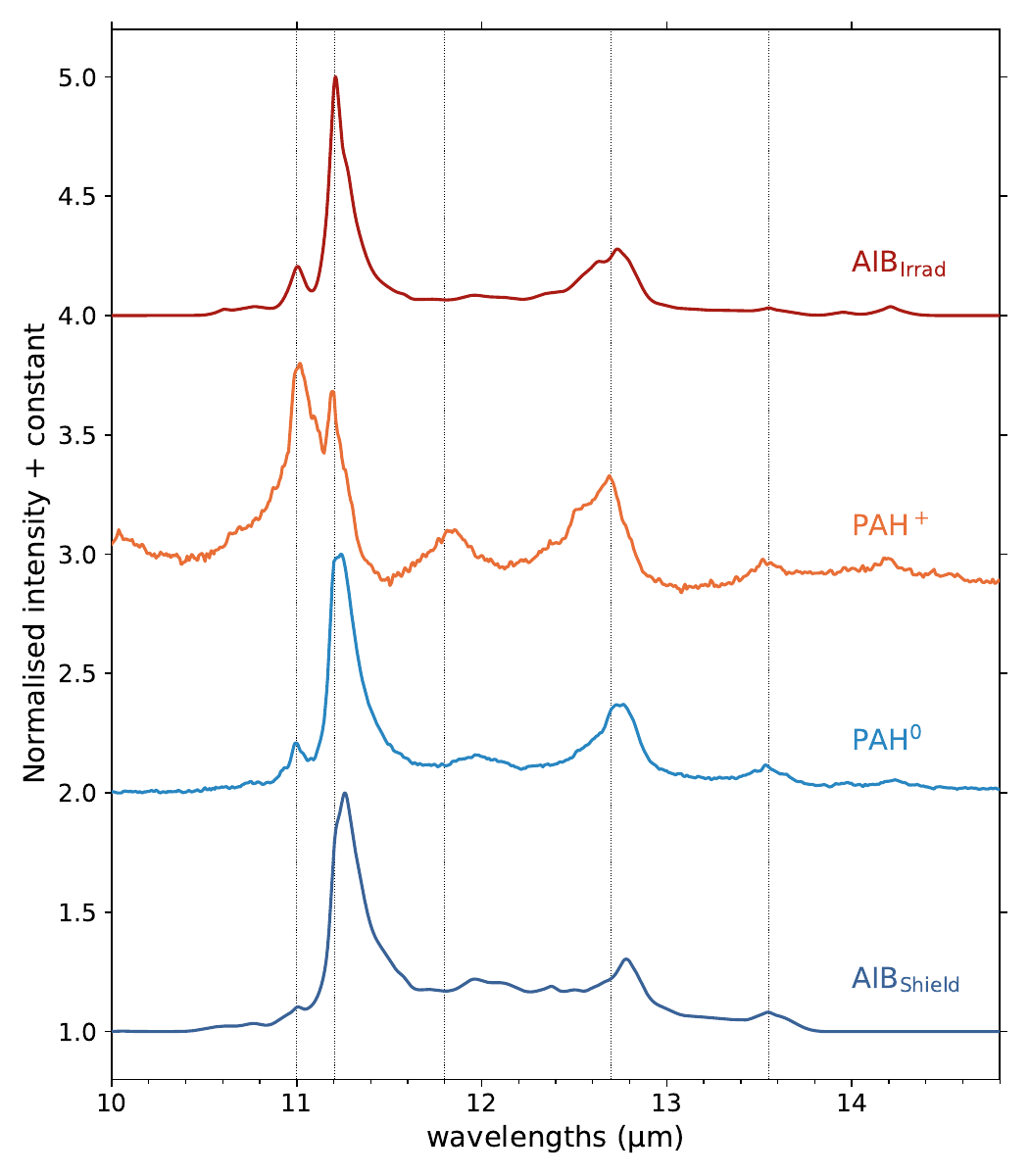}
    \caption{Elementary PAH spectra of cations and neutrals extracted from Spitzer observations of NGC 7023-NW by \citet{Rosenberg2011}, as well as \AIBirrad~and \AIBshield. The dotted vertical lines are placed at 11.0, 11.207, 11.8, 12.7 and 13.55~\textmu{}m.  }
    \label{fig:rosenberg_template}
\end{figure}

Fig.~\ref{fig:rosenberg_template} shows the template spectra of PAH cations and neutral extracted by \citet{Rosenberg2011} using Spitzer-IRS observations of the reflection nebulae NGC 7023-NW. \AIBshield~and \AIBirrad~ are also shown in the 10-15~\textmu{}m wavelengths range for comparison.
$\rm PAH^+$ shows strong 11.0, 11.2 and 12.7~\textmu{}m bands. $\rm PAH^0$ is dominated by the 11.2~\textmu{}m, which peaks around 11.25~\textmu{}m, and presents a moderately strong 12.7~\textmu{}m band. The presence of the 11.0~\textmu{}m feature in $\rm PAH^0$ was interpreted as an artifact caused by the limitations of unmixing. 
Spectrally, $\rm PAH^0$ is really close to \AIBshield. The spectral features of $\rm PAH^+$ are also consistent with that of \AIBirrad, with the presence of the 11.0~\textmu{}m band (though much stronger in $\rm PAH^+$), the 11.2~\textmu{}m peaking at 11.207~\textmu{}m and the 12.7~\textmu{}m band peaking at shorter wavelengths than in $\rm PAH^0$ and \AIBshield. 
One focus of the work of \citet{Rosenberg2011} was to compare the template spectra obtained with BSS with the NASA Ames PAH IR Spectroscopic Database. This enabled direct comparison with theoretical PAH spectra of different sizes and charges, and confirmed the nature of the carriers of $\rm PAH^+$, $\rm PAH^0$.

Following this approach, we fitted our template spectra using the pyPAHdb\footnote{\url{https://github.com/PAHdb/pyPAHdb}} spectral modeling tool \citep[][]{Shannon_pyPAHdb}.
The results are presented in Table~\ref{tab:pyPAHdb_results}, and the fits are shown in Fig.~\ref{fig:PAHdb_fit}.
These results indicate that \AIBirrad~corresponds to the emission of a population of mainly cationic species with an average size of $\rm N_C\approx73$, and a rather large fraction of small ($\rm N_C<50$) and medium ($\rm N_C\approx50-70$) PAHs (36\% for both). However, we note that the fit of \AIBirrad~shown in Fig.~\ref{fig:PAHdb_fit} shows large residuals, in particular for the 11.3~\textmu{}m band. 
For \AIBshield, the fit indicates a population which is predominantly neutral, with a high fraction (42\%) of small PAHs, hence an average PAH size of $\rm N_C\approx60$.

\begin{table}[h]
\caption{pyPAHdb best-fit parameters for \AIBirrad~and \AIBshield.}
\centering
\begin{tabular}{ c c c}
\hline\hline
 & \AIBirrad~& \AIBshield~\\ \hline

$f_{ \rm neutral}$  & 0.29 & 0.80 \\
$f_{ \rm cation}$  & 0.58 & 0.15 \\ 
$f_{ \rm anion}$ & 0.13 &  0.05 \\ \hline
$f_{ \rm large}$  & 0.29 & 0.36 \\
$f_{ \rm medium}$  & 0.36 & 0.22 \\
$f_{ \rm small}$  & 0.36 & 0.42 \\ \hline
$\rm N_C$ & 73 & 60 \\
\hline
\hline
\end{tabular}
\tablefoot{From top to bottom: the fraction of neutral, cationic and anionic PAHs, the fraction of large, medium and small PAHs, and the mean PAH size.}
\label{tab:pyPAHdb_results}
\end{table}

\subsection{Comparison of the spatial distribution of AIB emission in the Orion Bar from different approaches}\label{sect:PDRs4All}

As discussed in Section~\ref{sect:spatial_distribution}, the spatial distributions of \AIBshield~and \AIBirrad~in the NIRSpec observations (see the Fig.~2 of \citealt{schroetter_pdrs4all_2024}) are consistent with those presented for \AIBshield~and \AIBirrad~in Fig.~\ref{fig:MIRImap} for the counterpart MIRI observations.
However, the respective contributions of \AIBirrad~and \AIBshield~within the Orion Bar differ between MIRI and NIRSpec. 
More specifically, \AIBshield~contributes up to nearly 100\% of the total emission in the molecular PDR in NIRSpec, but only up to 50\% of the total emission in MIRI (cf. Fig.~\ref{fig:MIRImap}). 
This discrepancy could be due to the different wavelengths considered, which trace different PAH populations. The 3.3~\textmu{}m band (arising from aromatic C-H stretching modes) and the 3.4~\textmu{}m band covered by NIRSpec are dominated by the emission of small PAHs, with $\rm N_C<30$ \citep[e.g.][]{Tielens2005_book}. The NIRSpec data thus trace only a part of the PAH population, which is more sensitive to UV. 

\citet{pasquini_pdrs4all_2024} used bisecting k-means clustering to analyze the spectral variations of the 3.3, 6.2, 7.7-8.6 and 11.2~\textmu{}m bands across the Bar.
The variations of these bands appear directly related to the position in the mosaic, i.e. to the transition between the \ion{H}{ii} region, atomic PDR and DF1, DF2, DF3 (cf. Fig.~\ref{fig:NIRCam_MIRI_fov}), as first observed by \citet{chown_pdrs4all_2024, peeters_pdrs4all_2024}.
This pattern is reflected by individual band properties: the 6.2 and 7.7~\textmu{}m bands broaden when moving from the atomic to the molecular PDR, matching the increasing contribution of \AIBshield~(Fig.~\ref{fig:MIRImap}) across the same transition. 
The 8.6~\textmu{}m band's strength and peak position shift in step with \AIBirrad, which is strongest in the atomic PDR and weakens toward the molecular PDR. Similarly, the 11.0~\textmu{}m, which is carried by \AIBirrad, decreases (relatively to the 11.2~\textmu{}m band) when moving from the atomic to the molecular PDR.
Conversely, the 11.25~\textmu{}m feature, attributed to \AIBshield~(Fig.~\ref{fig:MIRImap}), shows the opposite behavior, i.e. peaking in the molecular PDR and declining toward the atomic PDR.
Together, these results show that the spatial distribution of \AIBshield~and \AIBirrad~accounts for the spectral variations observed for the 6.2, 7.7--8.6, and 11.2~\textmu{}m bands seen by \citet{pasquini_pdrs4all_2024} across the Bar.

\citet{khan_pdrs4all_2025} performed a detailed analysis of the C-H OOP bending modes in the 10--15~\textmu{}m region in the MIRI-MRS observations of the Orion Bar. 
Their study of the profile variations of the 11.2~\textmu{}m band revealed the distinct contributions from the 11.207~\textmu{}m and 11.25~\textmu{}m subfeatures, attributed to C-H OOP modes of solo hydrogen and VSGs, respectively. 
The spatial distributions of these two features (their Fig.~4) match those of \AIBirrad~and \AIBshield~(Fig.~\ref{fig:MIRImap}): the 11.207~\textmu{}m feature traces the atomic PDR, while the 11.25~\textmu{}m feature peaks near DF2 and DF3. 
\citet{khan_pdrs4all_2025} also found that the 11.0~\textmu{}m feature strongly correlates with the 11.207 component.
This is consistent with the fact that \AIBirrad~exhibits strong 11.005~\textmu{}m and 11.207~\textmu{}m components, while those are weak in \AIBshield~(cf. Fig.~\ref{fig:template_zoom}).

The variations of the 6-9~\textmu{}m wavelength range across the Orion Bar were analyzed by \citet{Khan2026_PDRs4All}. They observed an increase of the 6.2/8.6 and 7.7/8.6 ratios from the atomic to the molecular PDR, peaking at DF3. These variations are anti-correlated with the \AIBirrad~map of relative contribution, which peaks in the atomic PDR, declines toward the molecular PDR, and reaches a minimum at DF3. 
Since the 6.2/8.6 and 7.7/8.6 ratios are larger in \AIBshield~(cf. Tab~\ref{tab:ratios}), this anti-correlation is expected. This further confirms that a change in relative contribution of \AIBshield~and \AIBirrad~is consistent with the observed variations of the bands across the Bar.  \citet{Khan2026_PDRs4All} interpreted these variations as an effect of the photo-destruction of small PAHs with increasing FUV radiation.

\section{Chemical assignments of \AIBshield~and \AIBirrad~and the connection to PAH evolution in PDRs}\label{sect:chemical_prop}

The spectral characteristics presented in Section~\ref{sect:spectral_prop} and their link with earlier studies (Section~\ref{sect:previous_studies}) suggest a higher ionization fraction for the carriers of \AIBirrad~(higher 6.2, 7.7, 8.6/11.2 ratio, similarities with $\rm PAH^+$, comparison with PAHdb and spatial distribution). 
\citet{maragkoudakis_pdrs4all_2026} derived, using pyPAHdb, a cation fraction of $\approx 60\%$ in the atomic PDR, where the emission is dominated by \AIBirrad. This is in line with the cation fraction of 58\% (cf. Table~\ref{tab:pyPAHdb_results}) derived using pyPAHdb on \AIBirrad~directly. 
This confirms that the population carrying \AIBirrad~is dominated by cationic PAHs. This is also suggested by the presence of the 11.0~\textmu{}m feature in this spectrum, which is attributed to PAH cation \citep{Rosenberg2011}.
On the contrary, \AIBshield~emission peaks in the molecular PDR, where these latter authors found a cation fraction between 45--50\%. Given that \AIBirrad~and \AIBshield~contribute almost equally in this region, and \AIBirrad~is dominated by cationic PAHs by up to 60\%, this implies that \AIBshield~is instead dominated by neutral PAHs. This also consistent with the neutral fraction of 80\% presented Table~\ref{tab:pyPAHdb_results}.   
A similar analysis can be used to assess the size characteristics of the two populations: \citet{maragkoudakis_pdrs4all_2026} show that the
 mean PAH size in the atomic PDR, where \AIBirrad~dominates the emission, is $\rm N_C\approx 70$, while it is $\rm N_C\approx 60$ in the molecular PDR, where \AIBshield~dominates. These values are consistent with the ones presented Table~\ref{tab:pyPAHdb_results}. 
 Moreover, the broader 6.2 and 7.7~\textmu{}m bands in \AIBshield~are consistent with the emission from smaller PAHs, that can reach higher excitation temperatures, resulting in stronger anharmonic effects. This supports the interpretation that the average carriers of \AIBirrad~are larger than those responsible for \AIBshield. 
The very rich 10--15~\textmu{}m region in \AIBshield~indicates the presence of C-H OOP modes of solo, duo, trio, and quartet hydrogen, suggesting irregular molecules with a wide variety of shapes. 
The presence of the 11.25~\textmu{}m band in \AIBshield~can indicate that this template spectrum traces PAHs that are clustered \citep{khan_pdrs4all_2025}. This is also compatible with the presence of an extended red wing in the 11.2~\textmu{}m band, attributed to VSGs \citep{Rosenberg2011}.  
Lastly, in both the MIRI and NIRSpec observations \citep[see][]{schroetter_pdrs4all_2024} the \AIBshield~spectrum shows signatures from aliphatic bonds, i.e., at 3.4 and 6.8~\textmu{}m, respectively.

Overall, the spectral properties and spatial distributions of the two template spectra converge on a scenario in which:
\begin{itemize}
    \item The \AIBirrad~template spectrum is representative of the emission from a population of large and small, mainly cationic PAHs. This spectrum is predominantly produced in the atomic PDR. 
    \item The \AIBshield~template spectrum corresponds to a population made up of relatively smaller, neutral irregular PAHs containing aliphatic groups, and PAH clusters. This spectrum is mainly produced in the molecular PDR. 
\end{itemize}

The spatial distributions of \AIBshield~and \AIBirrad~therefore reflect the chemical evolution of the PAH population, driven by the changing UV radiation field. 
In the more UV shielded regions of the Orion molecular cloud, PAHs form clusters of small irregular PAHs containing aliphatic bonds. With the formation of young  massive stars within the cloud and associated expansion of the \ion{H}{ii} region, the pristine material from the molecular cloud is exposed to UV radiation, which processes this previously shielded population, dissociating any clusters. The \AIBshield~spectrum corresponds to this freshly exposed PAH population, that is weakly irradiated by UV. Under the effect of UV photons, this population evolves: weak aliphatic bonds are dissociated, small PAHs are destroyed, and the large surviving molecules are ionized, forming the PAH population that emits \AIBirrad.

\section{Conclusion}

Applying MASS-NMF on the PDRs4All MIRI-MRS observation of the Orion Bar, we extracted two AIB template spectra, namely \AIBirrad~and \AIBshield. 
These two spectra have distinct spectral properties, illustrated in Fig.~\ref{fig:template_zoom}.
\AIBshield~is characterized by strong 5.25 and 5.75~\textmu{}m bands (with respect to the 6.2~\textmu{}m band), a broad 6.2~\textmu{}m band, aliphatic bands (i.e., at 6.8~\textmu{}m), a plethora of subfeatures in the 7.7~\textmu{}m complex, and a strong 11.2~\textmu{}m complex (relative to the 6.2 and 7.7~\textmu{}m bands) peaking at 11.25~\textmu{}m. 
\AIBirrad~has prominent 6.2 and 7.7~\textmu{}m bands with respect to the 11.2~\textmu{}m band. The 7.7~\textmu{}m band clearly exhibits the 7.6 and 7.8~\textmu{}m subfeatures, and has a strong 8.6~\textmu{}m band that peaks at a longer wavelength than \AIBshield. 
The 11.2~\textmu{}m band peaks at 11.207~\textmu{}m, and the overall spectrum is smoother than \AIBshield. 
Overall, the multiple subfeatures summarized by \citet{chown_pdrs4all_2024} (presented in Table~\ref{tab:inv}) are subdivided into the two template spectra, with some belonging to either one, and some to both. The template spectra thus reveal which of these subfeatures are associated with each other, even though those are very weak and subtle.

The spectral characteristics and spatial distributions of \AIBshield~and \AIBirrad~across the Orion Bar indicate that they originate from two characteristic PAH populations. We interpret \AIBirrad~as being carried by ``UV-processed'' PAHs, which are mainly large ($\rm N_C>70$), cationic PAHs, while \AIBshield~traces a population of PAHs more shielded from UV irradiation, containing mainly small, neutral and cationic PAHs with irregular shapes and multiple aliphatic groups. An other component of the \AIBshield~population is VSGs or PAH clusters, contributing in the regions that are less exposed to UV.
The distribution of these two spectra then traces the evolution of the PAH populations across the Bar as observed in previous works, with small PAHs, VSGs and aliphatics being destroyed in the atomic PDR due to the high UV intensity. Deeper into the cloud, in the molecular PDR, the UV radiation field is attenuated so that the population is a mixture of large and small PAHs with aliphatic side groups, and a significant contribution of VSGs.

These AIB template spectra will serve future studies that investigate the properties of PAHs in various environment where the AIBs are observed by JWST with MIRI-MRS. We notably expect to use these templates as a diagnostic of PAH properties in a large sample of externally irradiated protoplanetary disks in the Orion Nebula Cluster, observed as part of the DiscOri program (PID:7534, Schroetter et al. in prep) with MIRI-MRS.

\begin{acknowledgements}
    This work is based on observations made with the
NASA/ESA/CSA James Webb Space Telescope. The data were obtained from the
Mikulski Archive for Space Telescopes at the Space Telescope Science Institute,
which is operated by the Association of Universities for Research in Astronomy, Inc., under NASA contract NAS 5-03127 for JWST. These observations are
associated with program \#1288 (DOI: 10.17909/pg4c-1737). Support for program
\#1288 was provided by NASA through a grant from the Space Telescope Science
Institute, which is operated by the Association of Universities for Research in
Astronomy, Inc., under NASA contract NAS 5-03127. 

OB, IS are funded by the Centre National d'Etudes Spatiales (CNES) through the APR program.

C.B. is thankful for an appointment at NASA Ames Research Center through the San Jos\'{e} State University Research Foundation (80NSSC22M0107) and gratefully acknowledges support from the Internal Scientist Funding Model (ISFM) Laboratory Astrophysics Directed Work Package at NASA Ames.

AF thanks funding from the European Research Council (ERC) under the European Union’s Horizon Europe research and innovation programme ERC-AdG-2022 (SUL4LIFE GA No. 101096293) and  project PID2022-137980NB-I00 funded by the Spanish Ministry of Science and Innovation/State Agency of Research MCIN/AEI/10.13039/501100011033 and by “ERDF A way of making Europe”.

J.R.G. thanks the Spanish MCINN for funding support under grant PID2023-146667NB-I00

EP acknowledges support from the University of Western Ontario, the Canadian Space Agency (CSA, 22JWGO1-16), and the Natural Sciences and Engineering Research Council of Canada.

TO acknowledges the support by the Japan Society for the Promotion of Science (JSPS) KAKENHI Grant Number JP24K07087. 

\end{acknowledgements}

\bibliographystyle{aa} 
\bibliography{biblio} 

\begin{appendix}

\onecolumn
\section{Detection of PAH subfeatures}

\begin{longtable}{llllccl }
    \caption{Catalogue of AIB features detected in the two template \AIBirrad~and \AIBshield. }\label{tab:inv} \\
\hline\\[-8pt]
   
    \multicolumn{2}{c}{AIB} & \multicolumn{2}{c}{Components} & \multicolumn{2}{c}{Detection} & Description \\
    \mic & cm$^{-1}$        &  \mic & cm$^{-1}$              & \AIBirrad~& \AIBshield~&    \\    

    \\[-8pt]
    \hline 
    \endfirsthead
    \caption{continued.}\\
    \hline\\[-8pt]
    \multicolumn{2}{c}{AIB} & \multicolumn{2}{c}{Components} & \multicolumn{2}{c}{Detection}  & Description \\
    \mic & cm$^{-1}$        &  \mic & cm$^{-1}$ & \AIBirrad~& \AIBshield~   &                     \\     

    \\[-8pt]
    \hline
\endhead
\hline
\endfoot
& \\[-5pt]

 \multicolumn{7}{c}{ \textbf{5--6~\mic region}}\\[5pt]
  \textbf{5.25} & 1905 & & & &  & \multirow{5}{15em}{Combination of out-of-plane bending modes} \\
         & & 5.18 & 1931   & -  &  - &\\ 
         & & 5.236 & 1910  & s  &  s &\\ 
         & & 5.30  & 1887   & m  &  s &\\ 
   5.435 & 1840    & &  & -  &  w &\\ 
   5.535 & 1807  & &  &  w &  - &\\ 
   \textbf{5.75} & 1739 & & & & & \\
        & & 5.642 & 1772   & s  &  m & \\ 
        & & 5.699 & 1755   & s  &  s & \\ 
        & & 5.755 & 1738   & s  &  s & \\ 
  5.878 & 1701  & &  & s  &  s &\\ 

\hline\\[1pt]
 \multicolumn{7}{c}{\textbf{6.2~\mic complex} }\\[5pt] 
   6.024 & 1660  & &  & w  & m &\\ 
 \textbf{6.2 }  &  1613 & & & & & \multirow{5}{15em}{Pure aromatic C-C stretching modes}\\
        & &  6.07 & 1647   & w  &  m & \\ 
        & & 6.212 & 1610   & s  &  s &\\    
        & & 6.395 & 1564   & m  &  m & \\ 
        & & 6.50 & 1538   & m  &  w & \\ 
\hline\\[2pt]
 \multicolumn{7}{c}{\textbf{7.7~\mic band} }\\[5pt] 
 6.638 & 1506   & & & m  &  w &\\ 
 6.711 & 1490   & & & m & - &  \multirow{3}{15em}{Aliphatic C-H bending modes}\\ 
 6.85 & 1460   & & & w  &  s  & \\ 
 6.943 & 1440   & & &  - &  s & \\ 
  \textbf{7.7} & 1299 & & & & &  \multirow{8}{15em}{aromatic C-C stretch and C-H in plane bend}\\
        & & 7.05 & 1418   & w  &  m & \\ 
         & & 7.1 & 1408   & w  &  m & \\ 
         & & 7.24 & 1381   &   -      &  m & \\ 
         & & 7.43 & 1346   & w  &  s & \\ 
         & & 7.626 & 1311   & s  &  s & \\ 
         & & 7.743 & 1291   &  m &  m & \\ 
         & & 7.775 & 1286   & w  & - & \\ 
         & & 7.85 & 1274   &  s  &  s &\\ 

 8.223 & 1216 & &   &  w &  s &\\ 
 8.33 & 1200  & &  & m  & -  & \\ 
 \hline\\[2pt]
 \multicolumn{7}{c}{\textbf{8.6~\mic complex} }\\[5pt] 
 \textbf{8.6} & 1163 & & & & &  \multirow{4}{15em}{C-H in-plane bending and mixture of C-C stretching modes} \\
         & & 8.46 & 1182   &  - &  m &\\ 
         & & 8.54 & 1171   &   - &  s &\\ 
         & & 8.6 & 1163   &  s  &   m &\\ 
         & & 8.74 & 1144   &  - &  m & \\ 
         & & 8.89 & 1125   &  m  & w & \\ 
 \hline \\[2pt] 
 \multicolumn{7}{c}{\textbf{10-15~\mic region} }\\[5pt] 
 10.95 & 913  & &  &  - &  w &\\ 
 11.005 & 909  & &  & s  & w & \\ 
 \textbf{11.2} & 893 & & & && \\ 
         & & 11.207 & 892   & s & w &  \multirow{3}{15em}{C-H out-of-plane bending modes of solo hydrogen} \\ 
         & & 11.25 & 889   & w & m&  \\ 
         & & 11.275 & 887   & w & s & \\ 
 \textbf{12} &  833 & & & & & \\[5pt] 
         & & 11.955 & 836   & w  &m  & \multirow{9}{15em}{C-H out-of-plane bending modes of duo and trio hydrogen} \\ 
         & & 12.125 & 825   &  w  &m  & \\ 
 \textbf{12.7} & 787 & & & && \\
         & & 12.38 & 808   & w & m & \\ 
         & & 12.52 & 799   &  w & m & \\ 
         & & 12.625 & 792   & s & m &  \\ 
         & & 12.729 & 786   & s & - & \\ 
         & & 12.779 & 783   & m  & s &\\  
         & & 12.98 & 770   &  w & w &\\  
 \textbf{13.5} & 741 & & & & & \multirow{3}{15em}{C-H out-of-plane bending modes of quartet hydrogen} \\
     & & 13.50 & 738   & - & w &  \\ 
     & & 13.55 & 738   & w & m & \\ 
     & & 13.62 & 734   & - & w& \\      
13.95 & 717  & &  & w  & - &  \\ 
14.21 & 704  & &  &m & - &\\

\end{longtable}
\tablefoot{Columns: (1), (2) Wavelength and wavenumber of AIB; nominal AIB names are listed in boldface; (3), (4) Wavelength and wavenumber of component; (5)-(6) Detection of AIB in \AIBshield~and \AIBshield, respectively. The strong bands are indicated with "s", the medium band with "m" and the weak band with "w".
}

\twocolumn

\section{MASS-NMF decomposition with different $r$}\label{sect:NMF_test_r}

We aim to investigate the viability of considering more components for the MASS-NMF decomposition. 
The MASS-NMF decomposition is applied on $\textbf{X}_{\rm AIB}$ as presented in Section~\ref{sect:NMF},  considering 100 different initializations (i.e different matrices \textbf{W}) for $r=3$ and $r=4$.
As in section~\ref{sect:NMF}, we extract the  $r=3$ and $r=4$ template spectra considering the mean of the obtained matrices \textbf{S}.
The uncertainties on the template spectra are taken as the standard deviation $\sigma$ of the matrices \textbf{S}.

\subsection{MASS-NMF with $r=3$}

\begin{figure*}
    \centering
    \includegraphics[width=\linewidth]{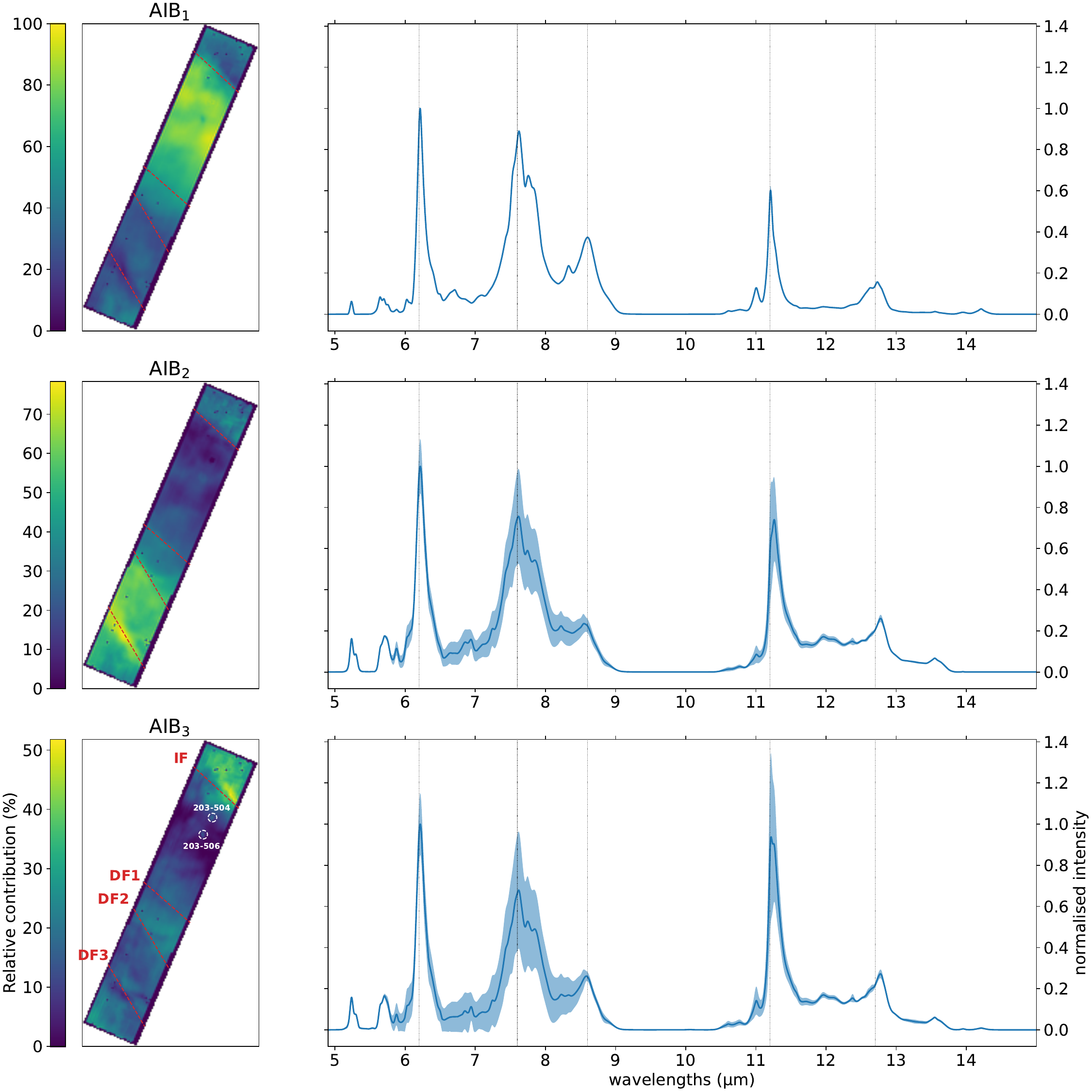}
    \caption{(right panels) Template spectra extracted from MASS-NMF considering $r=3$ components. 
    The blue curves correspond to the average spectra computed over the 100 iterations. The blue envelopes show the error estimated on the 100 iterations at 1-$\sigma$.
    (left panels) Relative spatial distribution of the different template spectra. }
    \label{fig:map_r3}
\end{figure*}

Fig.~\ref{fig:map_r3} shows the template spectra $\rm AIB_1$, $\rm AIB_2$ and $\rm AIB_3$, and their spatial distribution across the bar obtained considering $r=3$ components.  
Spatially, $\rm AIB_1$ mostly contributes to the emission in the atomic PDR, between IF and the DF1, similarly to \AIBirrad (cf. Fig.~\ref{fig:MIRImap}).  
In term of spectral properties, $\rm AIB_1$ shows great similarities with \AIBirrad, with strong 6.2 and 7.7~\textmu{}m band with respect to the 11.2~\textmu{}m band. 
$\rm AIB_2$ dominates the emission in the molecular PDR and peaks at DF3, and $\rm AIB_3$ dominates the emission in the \ion{H}{ii} region (which is in fact the emission from the background face-on PDR).
These two regions are dominated by the emission of \AIBshield~(cf. Fig.~\ref{fig:MIRImap}).
Moreover, $\rm AIB_2$ and $\rm AIB_3$ have spectral shapes which resemble that of \AIBshield, with strong 11.2~\textmu{}m complexes and large 6.2 and 7.7~\textmu{}m band. 
However, the large uncertainties on the 6.2, 7.7 and 11.2~\textmu{}m bands prevent the derivation of reliable 6.2/11.2 and 7.7/11.2 bands ratios, and indicates that the NMF is less stable upon initialization of the NMF with MASS for $r=3$.  

This overall suggests that considering $r=3$ is equivalent to splitting \AIBshield~into two subcomponents, but these subcomponents are unstable against initialization.
Furthermore, the uncertainties of $\rm AIB_2$ and $\rm AIB_3$ almost fully overlap, meaning these components are equal to within the margin of error.

\subsection{MASS-NMF with $r=4$}

\begin{figure*}
    \centering
    \includegraphics[width=0.95\linewidth]{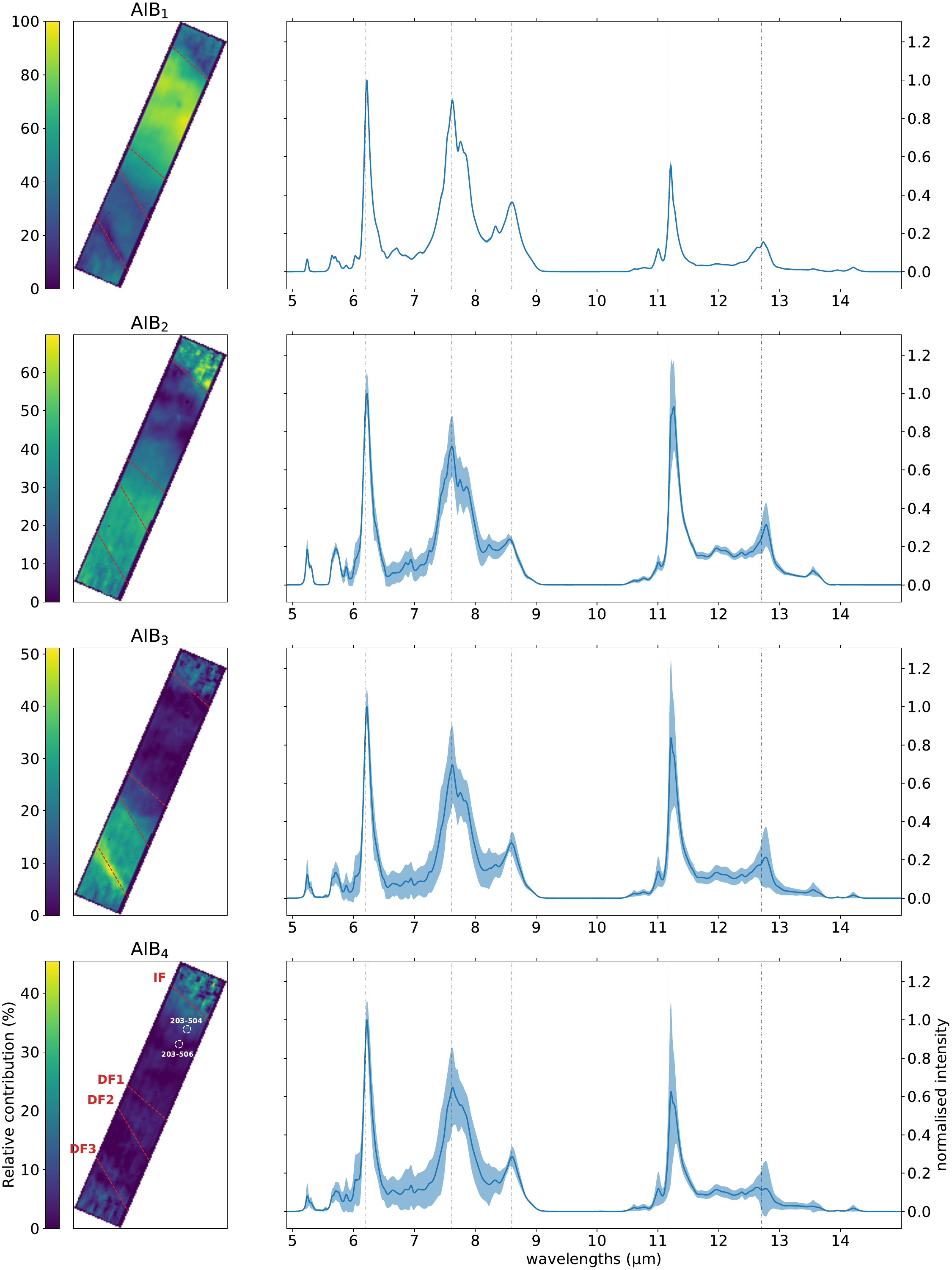}
    \caption{Same as Fig.~\ref{fig:map_r3} for $r=4$ components.}
    \label{fig:map_r4}
\end{figure*}

Fig.~\ref{fig:map_r4} shows the results of the NMF with $r=4$ components.
Again, $\rm AIB_1$ is consistent in term of spectral properties and spatial distribution with \AIBirrad. 
$\rm AIB_2$ mostly contributes in the \ion{H}{ii} region,  and moderately in the molecular PDR. 
The emission of $\rm AIB_3$ covers the molecular PDR and peaks at the location of DF3. 
$\rm AIB_4$ contributes to less than 10\% in most of the FOV. It peaks in the \ion{H}{ii} region, where it reaches $\approx 30\%$. 
Combined together, the spatial contributions of these three spectra  are consistent with that of \AIBshield. However, these spectra have great uncertainties on the main bands and so are unstable upon initialization of the NMF.

\subsection{NRMSE maps}

\begin{figure*}
    \includegraphics[width=\linewidth]{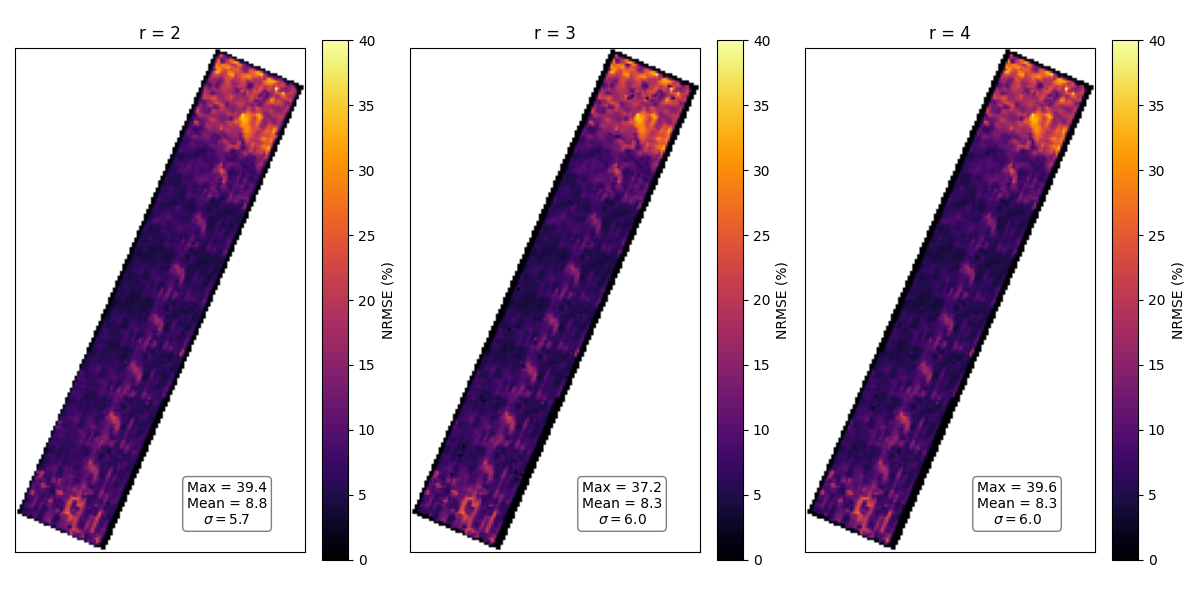}
    \caption{ Map of the total NRMSE of the fits performed using the template spectra extracted with $r=2$ (left panel), $r=3$ (middle panel) and $r=4$ (right panel). The maximum, mean and standard deviation measured across the NRMSE map are indicated in the bottom right corner of each panel. }
    \label{fig:map_NRMSE}
\end{figure*}

Fig.~\ref{fig:map_NRMSE} shows the NRMSE map of the fits using the template spectra extracted considering $r=2$, $r=3$ and $r=4$ components for the MASS-NMF (cf. section~\ref{sect:NMF}). 
The NRMSE is higher in the \ion{H}{ii} regions, where the emission is dominated by atomic lines. The three maps are really similarly.
The mean NRMSE still slightly decreases from 8.8 to 8.3\% when considering $r=3$ rather than $r=2$, at the cost of an increase of the standard deviation within the NRMSE map. 
For $r=4$, the mean NRMSE is also 8.3\%, but the peak value reaches 39.6\%.
Overall, these results indicate that considering $r>2$ does not significantly improve the quality of the fits, on top of introducing components that are unstable against initialization.


\section{Values of band ratios}
\begin{table}[h]
\caption{Peak to peak band ratio of a subset of AIBs for \AIBirrad~and \AIBshield.}
\centering
\begin{tabular}{ c c c}
\hline\hline
Ratio & \AIBirrad~& \AIBshield~\\ \hline
6.2/11.2 & 1.7 & 0.8\\
7.7/11.2 & 1.5 & 0.5 \\
8.6/11.2 & 0.6 & 0.2 \\ 
11.0/11.2 & 0.2 & 0.1 \\
12.8/11.2 & 0.3 & 0.3 \\\hline
6.2/8.6 & 2.8 & 5.6\\
7.7/8.6 & 2.5 & 3.2 \\

\hline

\hline

\end{tabular}
\label{tab:ratios}
\end{table}

\begin{figure*}
\section{Fit of \AIBirrad~and \AIBshield~with pyPAHdb}
    \centering
    \includegraphics[width=0.9\linewidth]{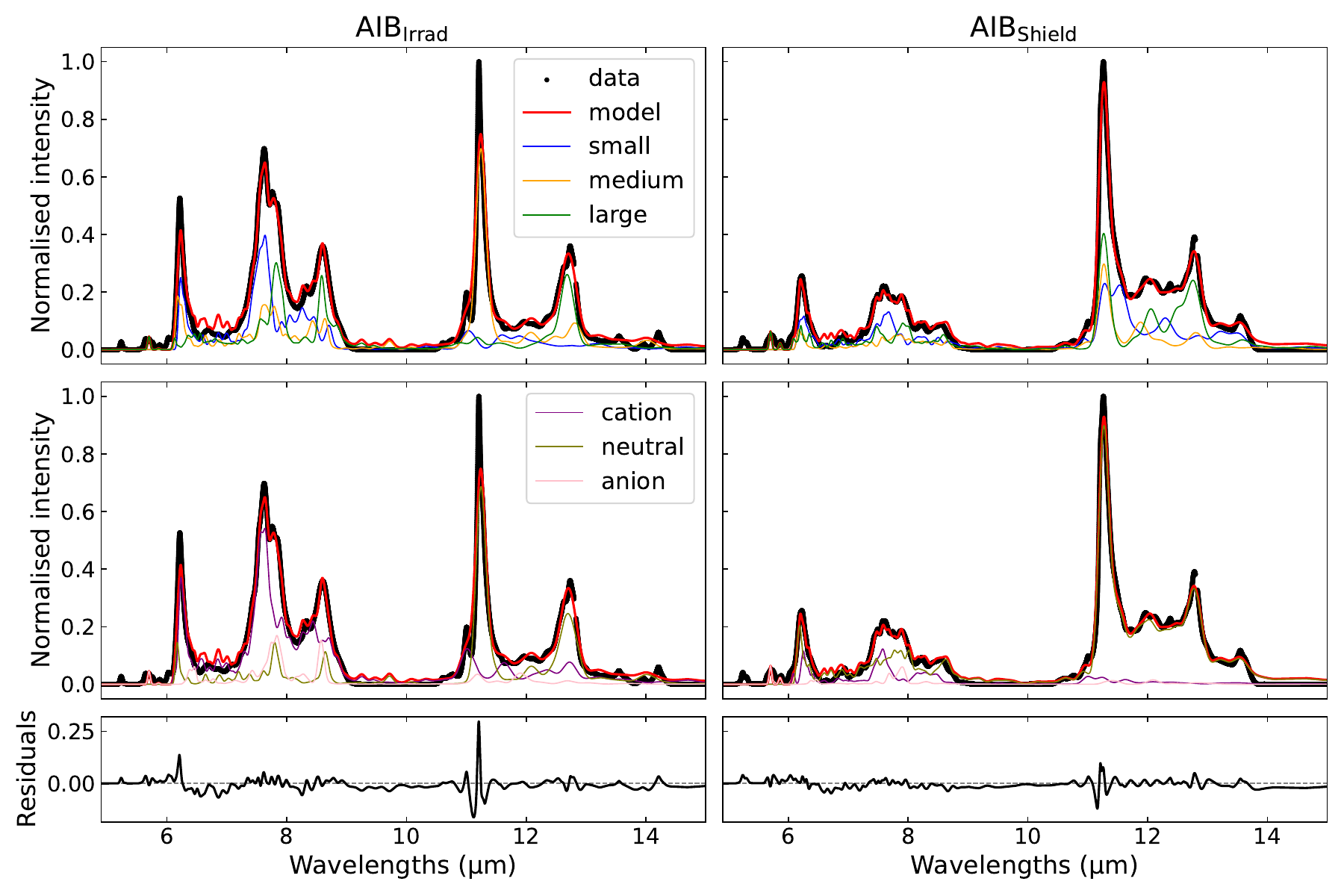}
    \caption{Fit of \AIBirrad ~(left) and \AIBshield~(right) with pyPAHdb. The data are shown with the black points, and the model are shown in red. The contributions of small ($\rm N_C<50$), medium ($\rm N_C\in50-70$) and large ($\rm N_C>70$) PAHs are shown in the top panels, while the contributions of PAH cations, neutrals and anions are shown in the middle panels. The residuals are indicated in the bottom panel of the respective template spectra.}
    \label{fig:PAHdb_fit}
\end{figure*}

\end{appendix}

\end{document}